\documentclass[prb,amsfonts,amssymb,floats,twocolumn,nolongbibliography,superscriptaddress,aps]{revtex4-2}
\usepackage[utf8]{inputenc}
\usepackage[T1]{fontenc}
\usepackage{amsmath}
\usepackage{float}
\usepackage[caption = false]{subfig}
\usepackage{color}
\usepackage{braket}
\usepackage{bm}
\usepackage{dsfont}
\usepackage{comment}
\usepackage{adjustbox}
\usepackage{graphicx}
\usepackage{enumitem}
\usepackage{bbm}
\usepackage{tikz}
\usetikzlibrary{patterns}

\usepackage[colorlinks=true]{hyperref}
\hypersetup{linkcolor=blue,citecolor=blue,urlcolor=blue}

\graphicspath{{./}{figs/}}

\begin{document}

\title{Symmetry-protected topological phases, conformal criticalities, and duality in exactly solvable SO($n$) spin chains}

\author{Sreejith Chulliparambil}
\thanks{The first two authors contributed equally.}
\affiliation{Institut f\"ur Theoretische Physik, Technische Universit\"at Dresden, 01062 Dresden, Germany}
\affiliation{Max-Planck-Institut f{\"u}r Physik komplexer Systeme, N{\"o}thnitzer Stra{\ss}e 38, 01187 Dresden, Germany}

\author{Hua-Chen Zhang}
\email{hczhang@phys.au.dk}
\affiliation{Department of Physics and Astronomy, Aarhus University, DK-8000 Aarhus C, Denmark}
\affiliation{Institut f\"ur Theoretische Physik, Technische Universit\"at Dresden, 01062 Dresden, Germany}

\author{Hong-Hao Tu}
\affiliation{Institut f\"ur Theoretische Physik, Technische Universit\"at Dresden, 01062 Dresden, Germany}

\date{\today}

\begin{abstract}
We introduce a family of SO($n$)-symmetric spin chains which generalize the transverse-field Ising chain for $n=1$. These spin chains are defined with Gamma matrices and can be exactly solved by mapping to $n$ species of itinerant Majorana fermions coupled to a static $\mathbb{Z}_2$ gauge field. Their phase diagrams include a critical point described by the $\mathrm{Spin}(n)_{1}$ conformal field theory as well as two distinct gapped phases. We show that one of the gapped phases is a trivial phase and the other realizes a symmetry-protected topological phase when $n \geq 2$. These two gapped phases are proved to be related to each other by a Kramers-Wannier duality. Furthermore, other elegant structures in the transverse-field Ising chain, such as the infinite-dimensional Onsager algebra, also carry over to our models.
\end{abstract}

\maketitle


\section{Introduction}
\label{sec:introduction}

Quantum spin chains have long been a fascinating area of research in physics, with some exactly solvable examples offering invaluable insights into many-body physics. One such model is the transverse-field Ising (TFI) chain, which is widely studied as an exemplary model for investigating quantum phase transitions and critical phenomena~\cite{cardy1996,sachdev2011,mussardo2020}. The TFI chain exhibits several intriguing properties from a theoretical perspective, such as the Kramers-Wannier duality~\cite{kramers1941} that maps between ordered and disordered phases, and the Onsager algebra~\cite{el-chaar2012}, which has played a key role in Onsager's epoch-making solution of the two-dimensional classical Ising model~\cite{onsager1944} (that is equivalent to the quantum TFI chain in certain limit~\cite{fradkin1978,kogut1979}) and ensures the integrability~\cite{dolan1982}.

While some of the TFI chain's desirable properties are retained in certain generalizations (such as $\mathbb{Z}_n$ clock chains~\cite{ortiz2012}), duality and other beautiful structures are more often only present in the low-energy limit, rather than at the lattice level. An example is the spin-1 bilinear-biquadratic chain near the Takhtajan-Babujian (TB) point~\cite{takhtajan1982,babujian1982}. The TB point is critical and its low-energy effective theory is the SU(2)$_2$ Wess-Zumino-Witten (WZW) model~\cite{affleck1987} (up to marginally irrelevant terms), which can be formulated in terms of three massless Majorana fermions~\cite{tsvelik1990,lecheminant2002}. The adjacent Haldane and dimerized phases are gapped, and the phase transition, occurring at the TB point, can be understood as a sign change of the Majorana fermion masses [where three masses are locked to be the same by the SO(3) symmetry]~\cite{tsvelik1990,lecheminant2002,liu2012,liu2014}. At the field theory level, this is similar to the situation in the TFI model where the transition between ordered and disordered phases is described by a sign change of a single Majorana fermion mass. However, the duality between Haldane and dimerized phases is no longer manifest on the lattice~\cite{chen2015}. A similar situation also arises in spin-1/2 ladders~\cite{shelton1996,nersesyan1997} and the SO($n$)-symmetric bilinear-biquadratic chain~\cite{tu2008a,tu2008b,tu2011,alet2011,okunishi2014}.

In this work, we present a solution of the TFI chain using the Majorana fermion representation of spin-1/2 Pauli operators~\cite{kitaev2006} and generalize it to derive a class of exactly solvable Gamma-matrix chains. This is largely inspired by a recent work~\cite{chulliparambil2020} generalizing the Kitaev's honeycomb model to exactly solvable Gamma-matrix models realizing the Kitaev's sixteenfold way of anyon theories in two dimensions. The Gamma-matrix chains, which we shall introduce in this work, possess an exact SO($n$) symmetry and can be represented by $n$ species of free itinerant Majorana fermions that are simultaneously coupled to a static $\mathbb{Z}_2$ gauge field. The phase diagram of these models includes a critical point described by the $\mathrm{Spin}(n)_{1}$ conformal field theory (CFT), as well as two distinct gapped phases. One of these two gapped phases is a symmetry-protected topological (SPT) phase (for $n \geq 2$), while the other is a trivial phase. The phase transition is indeed described by the sign change of $n$ Majorana fermion masses in the low-energy, long-wavelength limit and, at the same time, the appealing properties of the TFI chain, including the Kramers-Wannier duality and the Onsager algebra, are retained at the lattice level.

The rest of this paper is structured as follows. In Sec.~\ref{sec:models}, we introduce the exactly solvable SO($n$) spin chains and present their solutions. In Sec.~\ref{sec:criticality}, we show that the critical point of these SO($n$) spin chains is described by the $\mathrm{Spin}(n)_{1}$ CFT. In Sec.~\ref{sec:phases}, we turn to the two gapped phases separated by the $\mathrm{Spin}(n)_{1}$ critical point. The emphasis will be given to two limiting cases whose ground states are fixed-point wave functions with zero correlation length. In Sec.~\ref{sec:duality}, we show the presence of exact Kramers-Wannier duality and the Onsager algebra in our models. Sec.~\ref{sec:summary} summarizes this work and gives some outlook. Appendix~\ref{append:clifford-rep} provides explicit representations of the Gamma matrices forming the Clifford algebra $\mathrm{Cl}_{2n+1,0}(\mathbb{R})$.

\section{Models}
\label{sec:models}

\subsection{Transverse-field Ising chain}

We begin with the familiar spin-$1/2$ TFI chain, of which the Hamiltonian reads
\begin{equation}
\label{eq:TFIM-hamiltonian}
    H_{\textrm{TFI}} = - \sum_{j=1}^{N} \left( J \sigma^{z}_{j} \sigma^{z}_{j+1} - h \sigma^{x}_{j} \right),
\end{equation}
where $\sigma^{\alpha}$ ($\alpha=x,y,z$) are Pauli matrices and $J$, $h$ are real parameters. In~\eqref{eq:TFIM-hamiltonian} and throughout this work, we assume that $N$ is even and impose periodic boundary condition (e.g., $\sigma^{z}_{N+1} \equiv \sigma^{z}_{1}$ for the TFI chain). It is well-known that this model can be solved exactly by using the Jordan-Wigner transformation, under which~\eqref{eq:TFIM-hamiltonian} is mapped to a Hamiltonian of free fermions~\cite{pfeuty1970}. While the Jordan-Wigner transformation involves the introduction of nonlocal string operators, here we follow a different approach that is purely local and easily generalized to the $\mathrm{SO}(n)$ case, as we shall see below.

Our approach takes inspiration from the solution of the celebrated Kitaev's honeycomb model~\cite{kitaev2006}, where the Pauli matrices are represented by four Majorana operators $b^{x}$, $b^{y}$, $b^{z}$ and $c$ as follows:
\begin{equation}
\label{eq:pauli-parton}
    \sigma^{\alpha} = i b^{\alpha} c, \quad \alpha = x, y, z.
\end{equation}
Using the algebraic relations satisfied by the Majorana operators,
\begin{equation}
    \{ b^{\alpha}, b^{\beta} \} = 2 \delta_{\alpha \beta}, \quad \{ b^{\alpha}, c \} = 0, \quad c^{2} = 1,
\end{equation}
where $\alpha, \beta = x, y, z$, it is easy to verify that $\sigma^{x}$, $\sigma^{y}$, $\sigma^{z}$ defined by~\eqref{eq:pauli-parton} indeed satisfy the algebra expected for Pauli matrices. However, a constraint $b^{x} b^{y} b^{z} c = 1$ needs to be imposed to remove unphysical states, since four Majorana operators span a four-dimensional Hilbert space, whereas the (``physical'') Hilbert space of a spin-$1/2$ is two-dimensional.

\begin{figure}
\includegraphics[width=0.45\textwidth]{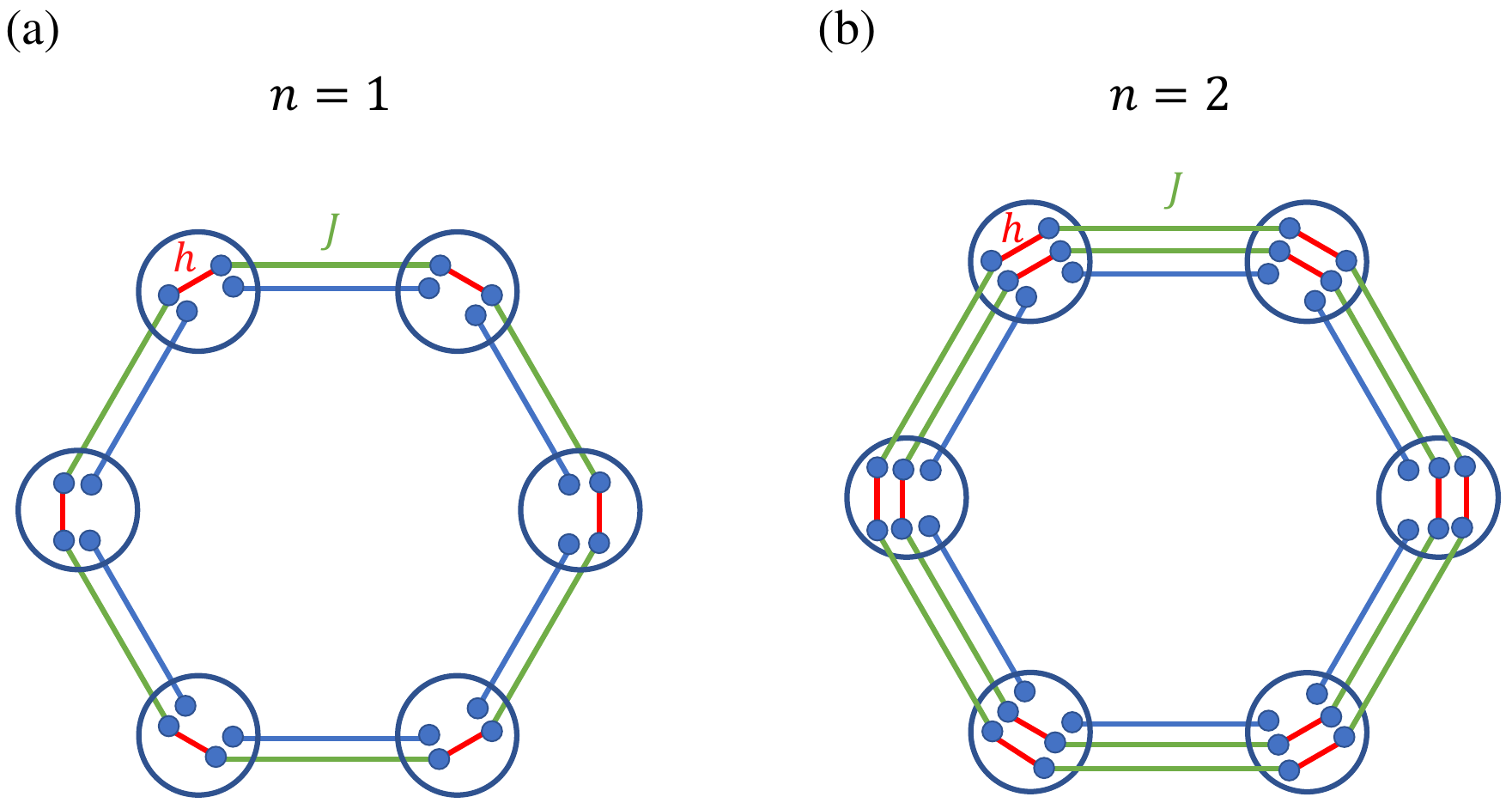}
\caption{Schematics of (a) the TFI chain ($n=1$) and (b) the $n=2$ model in the Majorana representation. The blue links represent the Majorana fermions from two neighboring sites which form the static $\mathbb{Z}_2$ gauge field. The green and red links represent the couplings of itinerant Majorana fermions at two neighboring sites and the same site, respectively. }
\label{fig:model}
\end{figure}

In the physical subspace defined by $b^{x} b^{y} b^{z} c = 1$, one can also represent the Pauli operators as $\sigma^{z} = -i b^{x} b^{y}$ and $\sigma^{x} = -i b^{y} b^{z}$. For the TFI Hamiltonian~\eqref{eq:TFIM-hamiltonian}, we use $\sigma^{x} = -i b^{y} b^{z}$ ($\sigma^{z} = i b^{z} c$) in the transverse-field (Ising coupling) term and obtain
\begin{equation}
\label{eq:TFIM-hamiltonian-parton}
    H_{\textrm{TFI}} = i \sum_{j=1}^{N} \left( J u_{j,j+1} b^{y}_{j} b^{z}_{j+1} + h b^{z}_{j} b^{y}_{j} \right)
\end{equation}
with $u_{j,j+1} \equiv i b^{x}_{j} c_{j+1}$, as depicted in Fig.~\ref{fig:model}(a). Obviously, the operators $u_{j,j+1}$ commute among themselves and $[H_{\textrm{TFI}},u_{j,j+1}] = 0$, and the eigenvalues of $u_{j,j+1}$ are $\pm 1$. Thus, $u_{j,j+1}$ can be interpreted as a static $\mathbb{Z}_{2}$ gauge field; the signs of $u_{j-1,j}$ and $u_{j,j+1}$ are flipped by the action of ``gauge transformation'' $D_{j} \equiv b^{x}_{j} b^{y}_{j} b^{z}_{j} c_{j}$, and the unphysical states are removed by the projector $\prod_{j} [(1 + D_{j})/2]$. The ``$\mathbb{Z}_{2}$ flux'' $w \equiv u_{1,2} u_{2,3} \cdots u_{N-1,N} u_{N,1}$, which is well-defined for a periodic chain, is gauge-invariant and has eigenvalues $\pm 1$ since $w^{2} = 1$. In fact, $w$ can be expressed in terms of the spin operators as $w = -\prod_{j=1}^{N} \sigma^{x}_{j}$ and is hence the global $\mathbb{Z}_{2}$ symmetry of the TFI chain. All configurations of the gauge field with the corresponding $w=+1$ (or $-1$) form an equivalent class modulo gauge transformations. Without loss of generality, we choose $u_{j,j+1} = u_{N,1} = 1~\text{with}~j = 1, \ldots, (N-1)$ as the representative of the class with $w = +1$, and $u_{j,j+1} = 1$ for $j = 1, \ldots, (N-1)$ and $u_{N,1} = -1$ as that with $w = -1$. In both cases, the Hamiltonian~\eqref{eq:TFIM-hamiltonian-parton} can be collectively written as
\begin{equation}
    H_{\textrm{TFI}} = i \sum_{j=1}^{N} \left( J b^{y}_{j} b^{z}_{j+1} + h b^{z}_{j} b^{y}_{j} \right)
\end{equation}
with the definition $c_{N+1} \equiv w c_{1}$. Because of the periodic boundary condition of spins ($\sigma^{z}_{N+1} \equiv \sigma^{z}_{1}$), the ``itinerant'' and ``gauge'' Majorana fermions ($b^{z}$ and $c$, respectively) must have the same boundary condition, which is periodic when $w = +1$ and antiperiodic when $w = -1$. We have seen that after this ``gauge-fixing'' procedure,~\eqref{eq:TFIM-hamiltonian} reduces to a Hamiltonian of free Majorana fermions (also known as the ``Kitaev chain''~\cite{kitaev2001}), which can readily be diagonalized by usual techniques of Fourier transformation.

\subsection{$\mathrm{SO}(n)$-symmetric spin chains}

Inspired by the Majorana representation~\eqref{eq:pauli-parton} of Pauli matrices, it is natural to define for a generic positive integer $n$ the following Majorana representation of $(2n+1)$ Gamma-matrices~\cite{wen2003b,yao2009,wu2009,ryu2009,chulliparambil2020,seifert2020,chulliparambil2021,jin2022,natori2023}:
\begin{equation}
    \Gamma^{\alpha} = ib^{\alpha}c, \quad \alpha = 1, 2, \ldots, (2n+1),
\end{equation}
where $b^{1}, \ldots, b^{2n+1}$ and $c$ are Majorana operators. These Gamma-matrices generate the Clifford algebra $\mathrm{Cl}_{2n+1,0}(\mathbb{R})$ with the anticommutation relation $\{ \Gamma^{\alpha}, \Gamma^{\beta} \} = 2\delta_{\alpha\beta}$. The (generalized) spin operators are represented in terms of the Gamma-matrices as well as their commutators $\Gamma^{\alpha\beta} \equiv \frac{i}{2} [\Gamma^{\alpha},\Gamma^{\beta}] = ib^{\alpha}b^{\beta}$, and the spin-chain Hamiltonian is defined as~\footnote{A caveat is in order. Given the close similarity between this Hamiltonian and that of the TFI chain~\eqref{eq:TFIM-hamiltonian}, it is tempting to interpret the coefficient of the $h$-term in Eq.~\eqref{eq:son-hamiltonian} as an external magnetic field. However, it is generally not the case except when $n=1$, for which the model reduces to the TFI chain itself; an example is given by the model with $n=2$, as detailed in Appendix~\ref{append:clifford-rep}. Instead, it is better to view the $J$- and $h$-terms as inter- and intra-site couplings, respectively. Moreover, the Hamiltonian~\eqref{eq:son-hamiltonian} exhibits the $\mathrm{SO}(n)$ symmetry for generic values of $J$ and $h$; this is not to be confused with the Hamiltonian given in~\eqref{eq:parent-hamiltonian-enhanced-symmetry}, which is another parent Hamiltonian of the ground state(s) when $h=0$ and is $\mathrm{SO}(n+1)$-symmetric.}
\begin{equation}
\label{eq:son-hamiltonian}
    H = \sum_{j=1}^{N} \sum_{\alpha = 1}^{n} \left( J \Gamma_{j}^{2\alpha-1,2n+1} \Gamma_{j+1}^{2\alpha} - h \Gamma_{j}^{2\alpha,2\alpha-1} \right).
\end{equation}
For the sake of definiteness, we make the assumption that $J>0,~h>0$ throughout this work. Similar to the case of the TFI chain, the dimension of the local Hilbert space spanned by the Majorana operators ($2^{n+1}$) has to be reduced by half to obtain that of the physical subspace ($2^{n}$); this can be achieved by restricting the local fermion parity
\begin{align}
\label{eq:local-fermion-parity}
    Q_{j} &= (ib^{1}_{j}b^{2}_{j}) \cdots (ib^{2n-1}_{j}b^{2n}_{j}) (ib^{2n+1}_{j}c_{j}) \nonumber \\
    &= i^{n} \Gamma^{1}_{j} \Gamma^{2}_{j} \cdots \Gamma^{2n+1}_{j}
\end{align}
to be $Q_j = +1$ or $Q_j = -1$ (note that $Q_{j}^{2} = 1$).

In the Majorana representation, the Hamiltonian~\eqref{eq:son-hamiltonian} is rewritten as
\begin{equation}
\label{eq:son-hamiltonian-parton}
    H = -i \sum_{j=1}^{N} \sum_{\alpha=1}^{n} \left( J u_{j,j+1} b^{2\alpha-1}_{j} b^{2\alpha}_{j+1} + h b^{2\alpha}_{j} b^{2\alpha-1}_{j} \right),
\end{equation}
where the example with $n=2$ is shown in Fig.~\ref{fig:model}(b). Here, $u_{j,j+1} \equiv i b^{2n+1}_{j} c_{j+1}$ is again a static $\mathbb{Z}_{2}$ gauge field whose eigenvalues $u_{j,j+1} = \pm 1$ label subspaces of the (extended) Hilbert space; the gauge-invariant ``loop'' operator
\begin{equation}
\label{eq:Z2-flux}
    w \equiv \left( \prod_{j=1}^{N-1} u_{j,j+1} \right) u_{N,1} = -\prod_{j=1}^{N} \Gamma^{2n+1}_{j}
\end{equation}
is a $\mathbb{Z}_2$ symmetry of the Hamiltonian~\eqref{eq:son-hamiltonian}. In fact, from the Majorana representation~\eqref{eq:son-hamiltonian-parton}, it is clear that when $n>1$ the Hamiltonian also exhibits a global $\mathrm{SO}(n)$ symmetry, which is associated with the ``rotation'' among Majorana operators $b^{1}_{j}, b^{3}_{j}, \ldots, b^{2n-1}_{j}$ as well as $b^{2}_{j}, b^{4}_{j}, \ldots, b^{2n}_{j}$. This symmetry transformation is generated by the operators $M^{\alpha\beta} \equiv \sum_{j=1}^{N} (\Gamma^{2\alpha-1,2\beta-1}_{j} + \Gamma^{2\alpha,2\beta}_{j}),~1 \leq \alpha < \beta \leq n$, which form the $\mathrm{so}(n)$ Lie algebra (up to a normalization constant) and commute with $w$ in~\eqref{eq:Z2-flux}. In Appendix~\ref{append:clifford-rep} it is shown that, by suitably choosing the Gamma-matrices generating $\mathrm{Cl}_{2n+1,0}(\mathbb{R})$, the model defined by~\eqref{eq:son-hamiltonian} is equivalent to the TFI chain~\eqref{eq:TFIM-hamiltonian} when $n = 1$ and reduces to a spin-$1/2$ bond-alternating XY chain when $n = 2$.

The total fermion parity of the whole chain is given by
\begin{equation}
    Q_{\textrm{total}} = \prod_{j=1}^{N} Q_{j} = Q_{\textrm{itinerant}} Q_{\textrm{gauge}},
\end{equation}
where
\begin{equation}
    Q_{\textrm{gauge}} = \prod_{j=1}^{N} \left( i b^{2n+1}_{j} c_{j} \right) = -w
\end{equation}
is the total parity of gauge fermions and
\begin{equation}
    Q_{\textrm{itinerant}} = \prod_{j=1}^{N} \left[ \left( i b^{1}_{j} b^{2}_{j} \right) \cdots \left( i b^{2n-1}_{j} b^{2n}_{j} \right) \right]
\end{equation}
is that of itinerant fermions. For either choice ($Q_j = +1$ or $Q_j = -1$) of the local fermion parity, we have $Q_{\textrm{total}}=1$ since $N$ is even. Thus, the parities of the itinerant and gauge fermions are bounded to be the same, $Q_{\textrm{gauge}} = Q_{\textrm{itinerant}} = -w$, which intimately relate to the boundary condition of Majorana fermions. Following the terminology used in CFT, we call the subspaces with $w = +1$ and $w = -1$ the Ramond (R) sector and Neveu-Schwarz (NS) sector, respectively. As we shall see in Sec.~\ref{sec:criticality}, projecting into sectors with definite eigenvalues of $Q_{\textrm{itinerant}}$ is crucial for revealing the nature of the quantum criticality.

\section{$\mathrm{Spin}(n)_{1}$ criticality}
\label{sec:criticality}

As in the case of TFI chain, the Hamiltonian~\eqref{eq:son-hamiltonian-parton} is bilinear in the fermion operators after the ``gauge fixing'':
\begin{equation}
\label{eq:son-hamiltonian-parton-gauge-fixed}
    H = -i \sum_{j=1}^{N} \sum_{\alpha=1}^{n} \left( J b^{2\alpha-1}_{j} b^{2\alpha}_{j+1} + h b^{2\alpha}_{j} b^{2\alpha-1}_{j} \right)
\end{equation}
for both the R sector (with $w = +1$) and the NS sector (with $w = -1$). To diagonalize this Hamiltonian and obtain the full energy spectrum, let us first perform an ``unfolding'' transformation by relabeling the Majorana operators as
\begin{eqnarray}
d^{\alpha}_{l} = \begin{cases}
\begin{array}{c}
b^{2\alpha}_{j}, \\
b^{2\alpha - 1}_{j},
\end{array} & \begin{array}{c}
l = 2j - 1 \\
l = 2j
\end{array} \end{cases},
\end{eqnarray}
in which $\alpha = 1, \ldots, n$ and $l = 1, 2, \ldots, 2N$; note that the length of the chain is doubled. We then proceed by transforming into Fourier space:
\begin{equation}
\label{eq:fourier-modes}
    d^{\alpha}_{l} = \frac{1}{\sqrt{N}} \sum_{k} \tilde{d}^{\alpha}_{k} e^{ikl},
\end{equation}
where the momenta are quantized according to the boundary condition $d^{\alpha}_{2N+1} \equiv w d^{\alpha}_{1}$ as
\begin{equation}
    k = \begin{cases}
        \pm \frac{\pi}{2N}, \pm \frac{3\pi}{2N}, \ldots, \pm \frac{(2N-1)\pi}{2N}, & \textrm{NS sector} \\
        0,\pm \frac{2\pi}{2N}, \pm \frac{4\pi}{2N}, \ldots, \pm \frac{(2N-2)\pi}{2N}, \pi, & \textrm{R sector}
    \end{cases}.
\end{equation}
With the inverse transformation
\begin{equation}
\label{eq:fourier-modes-2}
    \tilde{d}^{\alpha}_{k} = \frac{1}{2\sqrt{N}} \sum_{l=1}^{2N} d^{\alpha}_{l} e^{-ikl},
\end{equation}
it is straightforward to verify $\{ \tilde{d}^{\alpha}_{k}, (\tilde{d}^{\beta}_{k^{\prime}})^{\dagger} \} = \delta_{\alpha\beta} \delta_{kk^{\prime}}$ and $(\tilde{d}^{\alpha}_{k})^{\dagger} = \tilde{d}^{\alpha}_{-k}$ (note that modes with $k = \pi$ and $k = -\pi$ are identified). In particular, $\tilde{d}^{\alpha}_{0}$ and $\tilde{d}^{\alpha}_{\pi}$ are Hermitian and satisfy $(\tilde{d}^{\alpha}_{0})^{2} = (\tilde{d}^{\alpha}_{\pi})^{2} = 1/2$.

In terms of the Fourier modes~\eqref{eq:fourier-modes-2}, the Hamiltonian~\eqref{eq:son-hamiltonian-parton-gauge-fixed} is diagonalized via a unitary rotation in the space spanned by these modes. Let us postpone the discussion of the physics at generic $J$ and $h$ to Sec.~\ref{sec:phases} and focus in this Section on the case $J = h$, for which the Hamiltonian is already diagonalized without further rotation. Indeed, the Hamiltonian for both sectors now reads
\begin{equation}
\label{eq:hamiltonian-diagonalized}
    H^{\textrm{(NS/R)}} = \sum_{\alpha = 1}^{n} \sum_{k \in K^{\textrm{(NS/R)}}} \varepsilon(k) (\tilde{d}^{\alpha}_{k})^{\dagger} \tilde{d}^{\alpha}_{k} + E_{0}^{\textrm{(NS/R)}},
\end{equation}
where $K^{\textrm{(NS)}} \equiv \{ \frac{\pi}{2N}, \frac{3\pi}{2N}, \ldots, \frac{(2N-1)\pi}{2N} \},~K^{\textrm{(R)}} \equiv \{ \frac{2\pi}{2N}, \frac{4\pi}{2N}, \ldots, \frac{(2N-2)\pi}{2N} \}$, the dispersion $\varepsilon(k) = 4J \sin k$, and the ``vacuum'' energies in both sectors are
\begin{equation}
\label{eq:vacuum-energies}
    E_{0}^{\textrm{(NS)}} = -\frac{2nJ}{\sin{\frac{\pi}{2N}}}, \quad E_{0}^{\textrm{(R)}} = -2nJ \cot{\frac{\pi}{2N}}.
\end{equation}
It is worth emphasizing that the modes $\tilde{d}^{\alpha}_{0}$ and $\tilde{d}^{\alpha}_{\pi}$ in the R sector do not appear in~\eqref{eq:hamiltonian-diagonalized}; one could combine them as $f^{\alpha}_{0} \equiv (\tilde{d}^{\alpha}_{0} - i \tilde{d}^{\alpha}_{\pi})/\sqrt{2}$ to obtain $n$ fermionic ``zero modes''.

The complete thermodynamic properties of these models at temperature $T$ are encoded in the partition function, which is defined as $Z(T) = \mathrm{Tr}(e^{-H/T})$ (the Boltzmann constant is set to unity). In the path integral picture, this amounts to considering a periodic evolution of the quantum Hamiltonian with ``imaginary time'' $1/T$, resulting in a torus in the spacetime. According to the discussions above, the full partition function is the sum of contributions in the NS and R sectors, $Z = Z_{\textrm{NS}} + Z_{\textrm{R}}$. In the definition of $Z_{\textrm{NS}}$, the trace is taken over all energy eigenstates in the NS sector. However, as discussed in Sec.~\ref{sec:models}, the parity of the itinerant fermions must be restricted to $Q_{\textrm{itinerant}} = +1$ (resp. $-1$) in the NS (resp. R) sector. Thus, each of the energy eigenstates in the NS sector is labeled by a configuration $\{ F^{\alpha}_{k} \}_{k \in K^{\textrm{(NS)}}}$, where $F^{\alpha}_{k} = 0, 1$ is the occupation number of the mode $(\tilde{d}^{\alpha}_{k})^{\dagger}$; the parity constraint for the itinerant fermions requires that $F^{\textrm{(NS)}} \equiv \sum_{\alpha = 1}^{n} \sum_{k \in K^{\textrm{(NS)}}} F^{\alpha}_{k}$ is even. As the energy of this configuration is $E^{\textrm{(NS)}}(\{ F^{\alpha}_{k} \}) = E^{\textrm{(NS)}}_{0} + \sum_{\alpha = 1}^{n} \sum_{k \in K^{\textrm{(NS)}}} F^{\alpha}_{k} \varepsilon(k)$, the partition function from the NS sector reads
\begin{subequations}
\label{eqs:partition-function-lattice}
\begin{align}
    Z_{\textrm{NS}} =& \sum_{\{ F^{\alpha}_{k} \}} \frac{1 + (-1)^{F^{\textrm{(NS)}}}}{2} e^{-E^{\textrm{(NS)}}(\{ F^{\alpha}_{k} \})/T} \nonumber \\
    =&~\frac{1}{2} e^{-E^{\textrm{(NS)}}_{0}/T} \Bigg[ \prod_{k \in K^{\textrm{(NS)}}} \left( 1 + e^{-\varepsilon(k)/T} \right)^{n} \nonumber \\
    &~+ \prod_{k \in K^{\textrm{(NS)}}} \left( 1 - e^{-\varepsilon(k)/T} \right)^{n} \Bigg].
\end{align}
The partition function from the R sector can be computed similarly, except that one needs to take the zero modes into account. Denoting the occupation number of the zero mode $(f^{\alpha}_{0})^{\dagger}$ as $F^{\alpha}_{0} = 0, 1$, the parity constraint for the itinerant fermions now requires that $F^{\textrm{(R)}} \equiv \sum_{\alpha = 1}^{n} (F^{\alpha}_{0} + \sum_{k \in K^{\textrm{(R)}}} F^{\alpha}_{k})$ is odd. The energy of a generic configuration in the R sector is $E^{\textrm{(R)}}(\{ F^{\alpha}_{k} \}) = E^{\textrm{(R)}}_{0} + \sum_{\alpha = 1}^{n} \sum_{k \in K^{\textrm{(R)}}} F^{\alpha}_{k} \varepsilon(k)$, and the partition function reads
\begin{align}
    Z_{\textrm{R}} =& \sum_{\{ F^{\alpha}_{k} \}} \frac{1 - (-1)^{F^{\textrm{(R)}}}}{2} e^{-E^{\textrm{(R)}}(\{ F^{\alpha}_{k} \})/T} \nonumber \\
    =&~2^{n-1} e^{-E^{\textrm{(R)}}_{0}/T} \prod_{k \in K^{\textrm{(R)}}} \left( 1 + e^{-\varepsilon(k)/T} \right)^{n}.
\end{align}
\end{subequations}

From the expression of dispersion relation obtained above, it is clear that all the $n$ branches of fermionic modes are gapless at the point $J = h$, which is therefore a quantum critical point of the model~\eqref{eq:son-hamiltonian}. To characterize the effective field theory governing the low-temperature regime above this critical point, let us consider the continuum limit, which is defined as the limit $N \rightarrow \infty$ and $a \rightarrow 0$ while keeping $L = Na$ constant; here, $a$ denotes the lattice spacing (before ``unfolding'') and $L$ the length of the chain. Apparently, the low-energy modes are those adjacent to the Fermi points $k = 0$ and $k = \pi$, at which the dispersion is linearized as $\varepsilon(k) \sim 4Jk$ and $\varepsilon(k) \sim 4J(\pi - k)$, respectively. Linear dispersions near the Fermi points indicate that the low-energy physics of the critical system is described by a CFT; the latter, as we shall see next, can be identified by studying the partition function contributed by these linearized modes in the continuum limit. Furthermore, the ``vacuum'' energies~\eqref{eq:vacuum-energies} can be expanded as
\begin{subequations}
\label{eqs:vacuum-energies-expanded}
\begin{equation}
    E_{0}^{\textrm{(NS)}} = -2nJ \left( \frac{2N}{\pi} + \frac{\pi}{12N} + \mathcal{O}(N^{-3}) \right),
\end{equation}
\begin{equation}
    E_{0}^{\textrm{(R)}} = -2nJ \left( \frac{2N}{\pi} - \frac{\pi}{6N} + \mathcal{O}(N^{-3}) \right),
\end{equation}
\end{subequations}
where the leading terms in $N$ are divergent in the continuum limit. These terms can be dropped, however, as they are the same for both sectors and can be absorbed by an overall shift in energy. Keeping only the subleading terms in~\eqref{eqs:vacuum-energies-expanded} and substituting into~\eqref{eqs:partition-function-lattice}, one finds that the partition functions in the continuum limit can be succinctly represented in terms of certain special functions:
\begin{equation}
    \widetilde{Z}_{\textrm{NS}} = \frac{\theta_{3}^{n}(\tau) + \theta_{4}^{n}(\tau)}{2\eta^{n}(\tau)}, \quad \widetilde{Z}_{\textrm{R}} = \frac{\theta_{2}^{n}(\tau)}{2\eta^{n}(\tau)},
\end{equation}
where $\tau \equiv iv/LT$ with $v \equiv 2Ja$ being the ``velocity'' of the gapless modes. Here, the special functions ($q \equiv e^{2\pi i \tau}$)
\begin{subequations}
    \begin{equation}
        \theta_{2}(\tau) \equiv 2q^{1/8}\prod_{r=1}^{\infty}(1-q^{r})(1+q^{r})^{2},
    \end{equation}
    \begin{equation}
        \theta_{3}(\tau) \equiv \prod_{r=1}^{\infty}(1-q^{r})(1+q^{r-1/2})^{2},
    \end{equation}
    \begin{equation}
        \theta_{4}(\tau) \equiv \prod_{r=1}^{\infty}(1-q^{r})(1-q^{r-1/2})^{2}
    \end{equation}
\end{subequations}
are known as (standard) Jacobi's theta functions~\cite{mumford2006} and
\begin{equation}
    \eta(\tau) \equiv q^{1/24} \prod_{r=1}^{\infty} (1 - q^{r})
\end{equation}
is the Dedekind eta function.

In the context of rational CFTs~\cite{moore1988,cardy1989}, the torus partition function is given by certain modular-invariant combination of a finite number of holomorphic and antiholomorphic characters associated with irreducible highest-weight representations of the underlying chiral algebra (e.g., the Virasoro algebra for minimal models~\cite{belavin1984,friedan1984} or affine Lie algebras for WZW models~\cite{witten1984}), which are in one-to-one correspondence with the primary fields $\{ \mathfrak{a} \}$ of the theory. It turns out that the above-computed partition function, $\widetilde{Z} = \widetilde{Z}_{\textrm{NS}} + \widetilde{Z}_{\textrm{R}} = \sum_{\nu=2,3,4}\frac{\theta_{\nu}^{n}(\tau)}{2\eta^{n}(\tau)}$, is nothing but the modular-invariant partition function of the $\mathrm{Spin}(n)_{1}$ rational CFT~\footnote{Note that there are different notations for this theory in the literature. For example, the same theory is referred to as the $\mathrm{SO}(n)_{1}$ WZW model in Ref.~\cite{francesco1997}; here, we follow the terminology of Ref.~\cite{seiberg2016} where the name $\mathrm{SO}(n)_{1}$ is reserved for the corresponding ``spin CFT'', which depends on a particular choice of spin structures (i.e., boundary conditions of the fermions along the cycles of the spacetime torus). The $\mathrm{Spin}(n)_{1}$ CFT, on the contrary, is non-spin as the spin structures are summed over in computing the partition function}. To make the connection clear, let us briefly recall some basic facts pertinent to this theory~\cite{francesco1997,zhang2022}. For the $\mathrm{Spin}(n)_{1}$ CFT with odd $n$, there are three primary fields, $\mathfrak{a} = \boldsymbol{1}, \boldsymbol{v}, \boldsymbol{s}$; for the case of even $n$, the theory has four primary fields, $\mathfrak{a} = \boldsymbol{1}, \boldsymbol{v}, \boldsymbol{s}_{+}, \boldsymbol{s}_{-}$. Here, $\boldsymbol{1}$ is the identity field and $\boldsymbol{v}$ belongs to the vector representation of the Lie algebra $\mathrm{so}(n)$, to which the corresponding characters read
\begin{subequations}
\label{eq:son-characters}
\begin{equation}
    \chi_{\boldsymbol{1}}(q) = \frac{\theta_{3}^{\frac{n}{2}}(\tau) + \theta_{4}^{\frac{n}{2}}(\tau)}{2\eta^{\frac{n}{2}}(\tau)},\quad \chi_{\boldsymbol{v}}(q) = \frac{\theta_{3}^{\frac{n}{2}}(\tau) - \theta_{4}^{\frac{n}{2}}(\tau)}{2\eta^{\frac{n}{2}}(\tau)};
\end{equation}
the primary field(s) $\boldsymbol{s}$ (or, $\boldsymbol{s}_{+}$ and $\boldsymbol{s}_{-}$), on the other hand, belongs to the spinor representation(s) of $\mathrm{so}(n)$:
\begin{align}
    &\chi_{\boldsymbol{s}}(q) = \frac{\theta_{2}^{\frac{n}{2}}(\tau)}{\sqrt{2}\eta^{\frac{n}{2}}(\tau)} \quad \textrm{(for odd $n$)} \nonumber \\
    &\textrm{or,}~\chi_{\boldsymbol{s}_{+}}(q) = \chi_{\boldsymbol{s}_{-}}(q) = \frac{\theta_{2}^{\frac{n}{2}}(\tau)}{2\eta^{\frac{n}{2}}(\tau)} \quad \textrm{(for even $n$)}.
\end{align}
\end{subequations}
We note that by choosing $n = 1$, the above results reduce to those of the Ising CFT. With~\eqref{eq:son-characters}, one can easily verify that $\widetilde{Z} = \sum_{\mathfrak{a}} \vert \chi_{\mathfrak{a}}(q) \vert^{2}$ for both cases. In particular, the representations associated with $\boldsymbol{1}$ and $\boldsymbol{v}$ (resp. $\boldsymbol{s}$ or, $\boldsymbol{s}_{+}$ and $\boldsymbol{s}_{-}$) reside in the NS (resp. R) sector of the whole Hilbert space. Remarkably, the degeneracy of the ``vacuum'' in the R sector resulting from the zero modes agrees with the prediction by representation theory. In the case of odd $n$, the dimension of the spinor representation is given by the coefficient of the leading term (that is, the term $q^{n/24}$) in the expansion of $\chi_{\boldsymbol{s}}(q)$ in powers of $q$, which is equal to $2^{(n-1)/2}$. Likewise, the dimension of each of the two spinor representations, $\boldsymbol{s}_{\pm}$, of $\mathrm{so}(n)$ with even $n$ is $2^{(n-2)/2}$. In both cases, the corresponding coefficient in $\vert \chi_{\boldsymbol{s}}(q) \vert^{2}$ or $\vert \chi_{\boldsymbol{s}_{+}}(q) \vert^{2} + \vert \chi_{\boldsymbol{s}_{-}}(q) \vert^{2}$ is $2^{n-1}$, in agreement with the number of different ways to fill an odd number of zero modes out of $n$ ones. The degeneracies at higher energy levels can similarly be checked against the representation theory.

To summarize, we have computed exactly the partition function of the spin chain~\eqref{eq:son-hamiltonian} at the point $J = h$, thus identifying the latter as a conformal critical point described by the $\mathrm{Spin}(n)_{1}$ CFT. We conclude this Section by noting that several different approaches to realizing lattice models with the $\mathrm{Spin}(n)_{1}$ CFT being the low-energy effective theory were proposed in the literature. In Refs.~\cite{maansson2013,lahtinen2014,lahtinen2015}, a series of solvable spin-$1/2$ chains with $\mathrm{Spin}(n)_{1}$ critical points were constructed by using the ``anyon condensation'' mechanism. It would be interesting to see whether a particular Gamma-matrix choice in our Hamiltonian~\eqref{eq:son-hamiltonian} can reproduce their models. Other examples include the Reshetikhin model~\cite{reshetikhin1983,reshetikhin1985} (a critical point in the $\mathrm{SO}(n)$ bilinear-biquadratic chain~\cite{tu2008a,tu2008b,tu2011}) and the $\mathrm{SO}(n)$ generalization~\cite{tu2013a} of the Haldane-Shastry model~\cite{haldane1988,shastry1988}.

\section{Gapped phases}
\label{sec:phases}

In this Section, we analyze the ground-state phase diagram of our model~\eqref{eq:son-hamiltonian}. We have seen in Sec.~\ref{sec:models} that the TFI chain is mapped to Kitaev's Majorana chain after gauge-fixing (see also Ref.~\cite{fendley2012} for the equivalence through Jordan-Wigner transformation), in which one finds a topological phase as well as a trivial one that corresponds to the ordered and disordered phases of the TFI chain, respectively. In fact, the form~\eqref{eq:son-hamiltonian-parton-gauge-fixed} makes it manifest that our models are generalizations of Kitaev's Majorana chain for $n=1$, where $J$ and $h$ are, respectively, the inter- and intra-site couplings between the Majorana fermions. As the only quantum phase transition occurs at the critical point with $h/J = 1$ considered in Sec.~\ref{sec:criticality}, the nature of the gapped phases on both sides of this transition can be revealed by considering two limiting cases $ h/J \rightarrow 0^{+}$ and $h/J \rightarrow \infty$. In what follows, we first conduct a simple semi-quantitative analysis of these limiting cases using the Majorana representation. Subsequently, we focus on the model with $h=0$ and derive the ground state(s) as certain fixed-point matrix product states (MPSs) with zero correlation length. We then proceed to construct $\mathrm{SO}(n+1)$-symmetric parent Hamiltonians for these states and briefly comment on the connection with the existing classification scheme of SPT phases.

\begin{figure}
\includegraphics[width=0.45\textwidth]{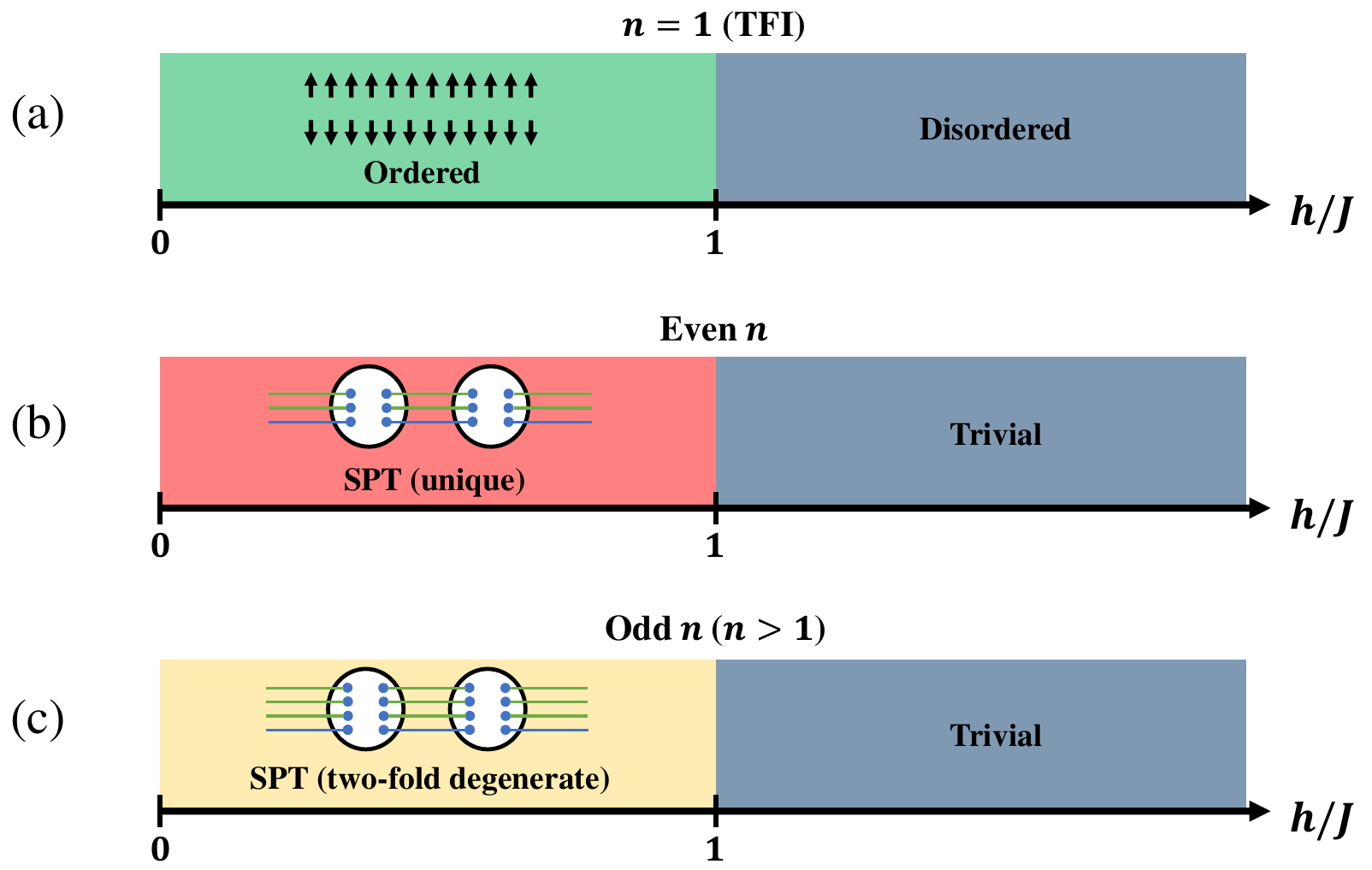}
\caption{Schematic ground-state phase diagrams of the Hamiltonian~\eqref{eq:son-hamiltonian}. The model with $n = 1$ reduces to (a) the TFI chain, where an ordered phase ($0 \leq h/J < 1$) and a disordered phase ($h/J > 1$) are separated by a critical point described by the Ising CFT. For models with $n>1$, the ordered and disordered phases are replaced by an SPT phase and a trivial one, respectively; the ground state in the SPT phase is unique for (b) the even $n$ case, whereas for (c) the odd $n$ case there are two-fold degenerate ground states arising from the spontaneous breaking of a $\mathbb{Z}_{2}$ symmetry. The critical point at $h/J = 1$ is described by the $\mathrm{Spin}(n)_{1}$ CFT for $n>1$.}
\label{fig:phasediagram}
\end{figure}

\subsection{Phase diagram}

Let us first look at the case $h/J \rightarrow \infty$. In this limit, the intra-site couplings dominate over the inter-site ones and the sites become independent of each other. One could recombine $b^{2\alpha-1}_{j}$ and $b^{2\alpha}_{j}$ to obtain a fermionic operator $f^{\alpha}_{j} = (b^{2\alpha}_{j} - i b^{2\alpha-1}_{j} )/2$ for $\alpha = 1, \ldots, n$, and the unique ground state is simply given by the fully empty state annihilated by all $f^{\alpha}_{j}$ (note that we have chosen $h>0$), which is also a product state in the spin basis. As the gap does not close for $h/J > 1$, we come to the conclusion that the system is in a trivial phase when $h/J > 1$.

More interesting is the case $0 \leq h/J < 1$. According to the representation~\eqref{eq:son-hamiltonian-parton} of the Hamiltonian in terms of Majorana fermions coupled to a static $\mathbb{Z}_{2}$ gauge field, the candidates for the ground state(s) admit the following form:
\begin{equation}
    \vert \Psi^{(\pm)} \rangle = P \vert \Psi_{\mathrm{F}}(\{ u^{\pm}_{0} \}) \rangle \otimes \vert \{ u^{\pm}_{0} \} \rangle.
\label{eq:projected-state}
\end{equation}
Here, $\{ u^{\pm}_{0} \}$ denotes an arbitrary configuration of the $\mathbb{Z}_{2}$ gauge field such that $w = \pm 1$ and $\vert \Psi_{\mathrm{F}}(\{ u^{\pm}_{0} \}) \rangle$ is the ground state of the Hamiltonian after gauge-fixing, which is quadratic in the itinerant Majorana fermions; $P$ is the projector enforcing the local parity constraint at each site, after the action of which the states become gauge-invariant. To survive projection, the total fermion parity of the unprojected state in~\eqref{eq:projected-state} must be even:
\begin{align}
\label{eq:global-parity-constraint}
    & \phantom{=} \; \vert \Psi_{\mathrm{F}}(\{ u^{\pm}_{0} \}) \rangle \otimes \vert \{ u^{\pm}_{0} \} \rangle  \nonumber \\
    &= Q_{\textrm{total}} \vert \Psi_{\mathrm{F}}(\{ u^{\pm}_{0} \}) \rangle \otimes \vert \{ u^{\pm}_{0} \} \rangle \nonumber \\
    &= Q_{\textrm{itinerant}} \vert \Psi_{\mathrm{F}}(\{ u^{\pm}_{0} \}) \rangle \otimes Q_{\textrm{gauge}} \vert \{ u^{\pm}_{0} \} \rangle \nonumber \\
    &= -w Q_{\textrm{itinerant}} \vert \Psi_{\mathrm{F}}(\{ u^{\pm}_{0} \}) \rangle \otimes \vert \{ u^{\pm}_{0} \} \rangle.
\end{align}
After the gauge-fixing, $n$ decoupled Kitaev chains of itinerant fermions admit the~\emph{same} dispersion, which in particular implies that they have the same fermion parity in $\vert \Psi_{\mathrm{F}}(\{ u^{\pm}_{0} \}) \rangle$. For even $n$, it follows immediately from~\eqref{eq:global-parity-constraint} that $w$ must be $-1$ (due to $Q_{\textrm{itinerant}}=+1$ in the ground state), i.e., the unique ground state resides in the NS sector. For odd $n$, on the contrary, both $w = +1$ and $w = -1$ are allowed by the parity constraint, which indicates a \emph{spontaneous} breaking of the global $\mathbb{Z}_{2}$ symmetry~\eqref{eq:Z2-flux}. In the latter case, a more detailed analysis shows that the lowest-energy states in both sectors are quasi-degenerate with an energy splitting that is exponentially small in the system size. Instead of carrying out this analysis explicitly, we briefly summarize the above results in Fig.~\ref{fig:phasediagram} and focus in the remainder of this Section on the model with $h = 0$, for which the degeneracy becomes exact.

\subsection{Fixed-point MPS and $\mathrm{SO}(n+1)$-symmetric parent Hamiltonian}

We shall see that for all $n \geq 2$, the ground states at $h = 0$ can be neatly expressed as fixed-point MPSs by ``splitting'' each physical site into two auxiliary ones (c.f. Fig.~\ref{fig:fixed-point_MPS} below). This is achieved using the fact that the representation of the Clifford algebra $\mathrm{Cl}_{2n+1,0}(\mathbb{R})$ with $n = 2k$ or $n = 2k - 1$ can be constructed utilizing two copies of that of $\mathrm{Cl}_{2k+1,0}(\mathbb{R})$ (see Appendix~\ref{append:clifford-rep}), which is generated by $\Lambda^{\alpha},~\alpha = 1, 2, \ldots, (2k+1)$. An explicit representation is given in~\eqref{eq:lambda-definition}, where the $\Lambda^{\alpha}$ are $2^{k}$-dimensional matrices. It can be readily verified that their commutators $\Lambda^{\alpha\beta} \equiv \frac{i}{2} [\Lambda^{\alpha},\Lambda^{\beta}]$ with $1 \leq \alpha < \beta \leq (n+1)$ generate the Lie algebra $\mathrm{so}(n+1)$; the representation of the latter induced by~\eqref{eq:lambda-definition} is nothing but the spinor representation that we mentioned in Sec.~\ref{sec:criticality}. For $n = 2k$, this representation is irreducible and denoted as $D^{k}$; for $n = 2k - 1$, however, it is reducible due to the existence of $\Lambda^{2k+1}$ that commutes with all the $\Lambda^{\alpha\beta}$ ($1 \leq \alpha < \beta \leq 2k$) and hence decomposes into two irreducible ones, which are denoted as $D^{k,+}$ and $D^{k,-}$~\footnote{To tally with the notation we used in Sec.~\ref{sec:criticality}, we note that the representation $D^{k}$ corresponds to the primary field $\boldsymbol{s}$ of the $\mathrm{Spin}(2k+1)_1$ CFT, and $D^{k,\pm}$ correspond to $\boldsymbol{s}_{\pm}$ of the $\mathrm{Spin}(2k)_1$ CFT}. In both cases, the Hamiltonian after this ``splitting'' reads
\begin{align}
    H_{h=0} &= J \sum_{j=1}^{N} \sum_{\alpha = 1}^{n} \Lambda_{2j}^{\alpha,n+1} \Lambda_{2j+1}^{\alpha,n+1} \nonumber \\
    &= \frac{J}{2} \sum_{j=1}^{N} \sum_{\alpha = 1}^{n} \left[ \left( \Lambda_{2j}^{\alpha,n+1} + \Lambda_{2j+1}^{\alpha,n+1} \right)^{2} - 2 \right],
\end{align}
which is a sum of mutually commuting local terms. As the bond between auxiliary sites $2j$ and $2j + 1$ is just the physical bond between sites $j$ and $j+1$ in the original model, the ground state(s) is given by the tensor product of the ``bond singlets'' that are annihilated by $( \Lambda_{2j}^{\alpha,n+1} + \Lambda_{2j+1}^{\alpha,n+1} )$. In fact, it is straightforward to verify that the $2^{k} \times 2^{k}$ matrix $R \equiv \sigma^{y}_{1} \otimes \sigma^{x}_{2} \otimes \sigma^{y}_{3} \otimes \sigma^{x}_{4} \otimes \cdots$ satisfies the relation $R^{-1} \Lambda^{\alpha,n+1} R = -(\Lambda^{\alpha,n+1})^{\mathrm{T}}$~\cite{georgi1999}; thus, the singlet between physical sites $j$ and $j+1$ is represented in an orthonormal basis $\{ \vert x \rangle \}_{x = 1}^{2^{k}}$ of the spinor representation as
\begin{equation}
\label{eq:so(n+1)-singlet}
    \vert (0) \rangle_{j,j+1} = \sum_{x,y = 1}^{2^{k}} R_{xy} \vert x \rangle_{2j} \vert y \rangle_{2j+1}.
\end{equation}
For $n = 2k$, the (unique) ground state is therefore given by $\vert \Psi_{h=0} \rangle = \prod_{j=1}^{N} \vert (0) \rangle_{j,j+1}$ [Fig.~\ref{fig:fixed-point_MPS}(a)]. Complications appear in the case $n = 2k - 1$ due to the reducibility of the corresponding spinor representation; here, the irreducible representation $D^{k,\pm}$ at auxiliary site $l$ is obtained by acting with the projector $P_{l}^{\pm} \equiv (1 \pm \Lambda_{l}^{2k+1})/2$. As one can easily show, these projectors are related to the local fermion parity by $P_{2j-1}^{+}P_{2j}^{+} + P_{2j-1}^{-}P_{2j}^{-} = [1 + (-1)^{k}Q_{j}]/2,~P_{2j-1}^{+}P_{2j}^{-} + P_{2j-1}^{-}P_{2j}^{+} = [1 - (-1)^{k}Q_{j}]/2$. By choosing the eigenvalue of $Q_{j}$ to be $+1$ at each physical site $j$ without loss of generality, this decomposition results in two degenerate ground states [Fig.~\ref{fig:fixed-point_MPS}(b)]:
\begin{subequations}
\begin{equation}
    \vert \Psi_{h=0}^{(\mathrm{\uppercase\expandafter{\romannumeral1}})} \rangle = \prod_{j=1}^{N} \left[ P_{2j}^{+}P_{2j+1}^{+} \vert (0) \rangle_{j,j+1} \right],
\end{equation}
\begin{equation}
    \vert \Psi_{h=0}^{(\mathrm{\uppercase\expandafter{\romannumeral2}})} \rangle = \prod_{j=1}^{N} \left[ P_{2j}^{-}P_{2j+1}^{-} \vert (0) \rangle_{j,j+1} \right]
\end{equation}
\end{subequations}
for even $k$, or [Fig.~\ref{fig:fixed-point_MPS}(c)]
\begin{subequations}
\begin{equation}
    \vert \Psi_{h=0}^{(\mathrm{\uppercase\expandafter{\romannumeral1}})} \rangle = \prod_{j=1}^{N} \left[ P_{2j}^{-}P_{2j+1}^{+} \vert (0) \rangle_{j,j+1} \right],
\end{equation}
\begin{equation}
    \vert \Psi_{h=0}^{(\mathrm{\uppercase\expandafter{\romannumeral2}})} \rangle = \prod_{j=1}^{N} \left[ P_{2j}^{+}P_{2j+1}^{-} \vert (0) \rangle_{j,j+1} \right]
\end{equation}
\end{subequations}
for odd $k$, in agreement with our semi-quantitative analysis using the Majorana representation.

\begin{figure}
\includegraphics[width=0.45\textwidth]{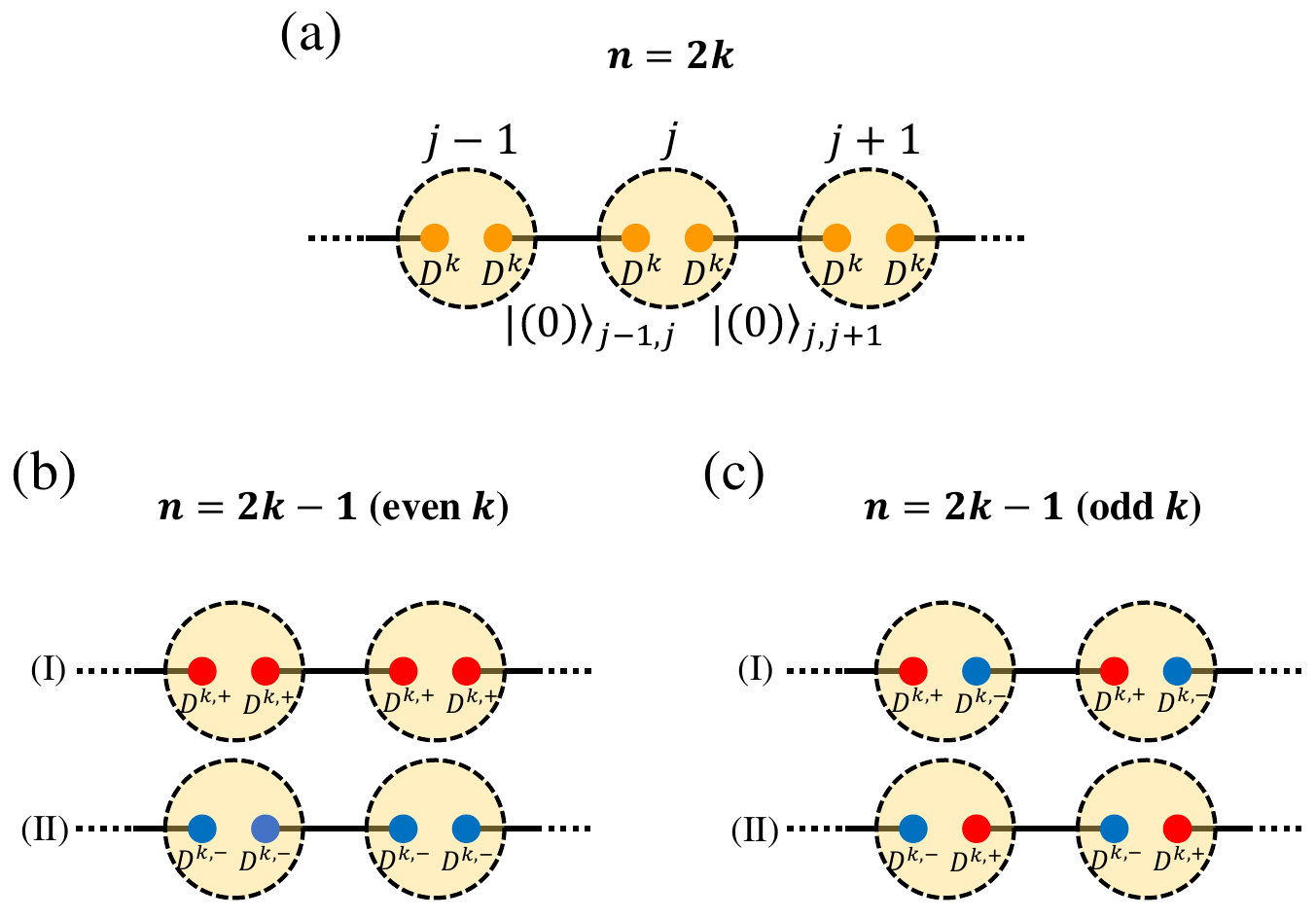}
\caption{Structure of the fixed-point MPSs. Each auxiliary site carries a Hilbert space that is given by an irreducible spinor representation of the Lie algebra $\mathrm{so}(n+1)$. The SPT phase has a unique ground state for (a) the $n = 2k$ case and two-fold degenerate ground states for the $n = 2k - 1$ case with (b) even $k$ and (c) odd $k$, respectively.}
\label{fig:fixed-point_MPS}
\end{figure}

At $h = 0$, the ``bond-singlet'' form of the ground state(s) indicates that the latter admits parent Hamiltonian with symmetry enhanced to $\mathrm{SO}(n+1)$. Indeed, as the $\mathrm{SO}(n+1)$-singlet given in~\eqref{eq:so(n+1)-singlet} satisfies $\Lambda^{\alpha\beta}_{2j}\Lambda^{\alpha\beta}_{2j+1} \vert (0) \rangle_{j,j+1} = - \vert (0) \rangle_{j,j+1},~1 \leq \alpha < \beta \leq (n+1)$, $\vert (0) \rangle_{j,j+1}$ minimizes the eigenvalue of an $\mathrm{SO}(n+1)$-symmetric Heisenberg interaction $K_{2j,2j+1} \equiv \sum_{1 \leq \alpha < \beta \leq (n+1)} \Lambda^{\alpha\beta}_{2j}\Lambda^{\alpha\beta}_{2j+1}$. Thus, a parent Hamiltonian of the ground state $\vert \Psi_{h=0} \rangle = \prod_{j=1}^{N} \vert (0) \rangle_{j,j+1}$ for $n = 2k$ is given by
\begin{equation}
\label{eq:parent-hamiltonian-enhanced-symmetry}
    \widetilde{H} = \sum_{j=1}^{N} K_{2j,2j+1} = \sum_{j=1}^{N} \sum_{1 \leq \alpha < \beta \leq (n+1)} \Gamma^{2\alpha-1,2\beta-1}_{j} \Gamma^{2\alpha,2\beta}_{j+1},
\end{equation}
where we have ``grouped'' two auxiliary sites back to get a physical one, and used the convention $\Gamma^{2\alpha,2n+2} \equiv \Gamma^{2\alpha}$. For $n = 2k - 1$, as discussed above, one can project into the subspace with $Q_{j} = +1,~j = 1, \ldots, N$, and conclude that $\vert \Psi_{h=0}^{(\mathrm{\uppercase\expandafter{\romannumeral1}})} \rangle$ and $\vert \Psi_{h=0}^{(\mathrm{\uppercase\expandafter{\romannumeral2}})} \rangle$ are degenerate ground states of $\widetilde{H}$.

Finally, let us remark on how the above results fit into the general framework for the classification of SPT phases. For one-dimensional bosonic systems such as the spin chains we are considering, the classification is done by using the group cohomology theory~\cite{gu2009,pollmann2010,chen2011,schuch2011}. More precisely, for Hamiltonians that commute with the action of a simple Lie algebra $\mathfrak{g}$, the SPT phases that the ground states belong to are in one-to-one correspondence with the elements of the cohomology group $H^{2}(G/G^{\prime}, \mathrm{U}(1)) \cong G^{\prime}$. Here, $G$ is the simply connected Lie group associated with $\mathfrak{g}$, and $G^{\prime}$ is the largest central subgroup of $G$ that acts trivially on the local Hilbert spaces. Evidently, $G^{\prime}$ depends on the specific representation of $\mathfrak{g}$ at each site. According to the discussions in the last paragraph, the relevant Lie algebra in our case is $\mathfrak{g} = \mathrm{so}(n+1)$ with the associated simply connected Lie group given by $G = \mathrm{Spin}(n+1)$. For our models, we shall show that $G^{\prime}$ is just the very largest central subgroup of $G$, that is, the center $\mathcal{Z}(G)$ itself. To this end, let us invoke the mathematical fact that each element of $H^{2}(G/\mathcal{Z}(G), \mathrm{U}(1))$, which is the projective class of a linear representation of $G$ (hence also a representation of $\mathfrak{g}$) with highest weight $\lambda$, is labeled by the congruence class $[\lambda]$ of $\lambda$. For the case $n = 2k$, it turns out that $\mathcal{Z}(G)$ is represented by $\rho_{\lambda} = (-1)^{[\lambda]}\mathbb{I}$ with $[\lambda]$ an element of $\mathbb{Z}_{2}$~\cite{duivenvoorden2013}. As the local Hilbert space at each physical site consists of the tensor product $D^{k} \otimes D^{k}$ [c.f. Fig.~\ref{fig:fixed-point_MPS}(a)], one can see that $\mathcal{Z}(G)$ acts trivially on it since $\rho_{D^{k}} \otimes \rho_{D^{k}} = \mathbb{I}$. The situation is slightly more complicated for $n = 2k - 1$, where $[\lambda] = [[\lambda]_{1}, [\lambda]_{2}]$ with $[\lambda]_{1} \in \mathbb{Z}_{2}$ and $[\lambda]_{2} \in \mathbb{Z}_{4}$; for the two irreducible spinor representations $D^{k,+}$ and $D^{k,-}$, the congruence classes are given by $[1,~k-2~(\mathrm{mod}~4)]$ and $[1,~k~(\mathrm{mod}~4)]$, respectively~\cite{ramond2010}. If $k$ is even, it is easily seen that these congruence classes form the group $\mathbb{Z}_{2} \times \mathbb{Z}_{2}$: $[1, 0] + [1, 0] = [1, 2] + [1, 2] = [0, 0],~[1, 0] + [1, 2] = [1, 2] + [1, 0] = [0, 2]$. In fact, it is known that the elements in $\mathcal{Z}(G)$ corresponding to the two $\mathbb{Z}_{2}$ subgroups are represented by $\rho_{\lambda}^{(1)} = (-1)^{[\lambda]_{1}}\mathbb{I}$ and $\rho_{\lambda}^{(2)} = e^{\frac{i\pi}{2}[\lambda]_{2}}\mathbb{I}$, respectively~\cite{duivenvoorden2013}; in particular, both of them act trivially on $D^{k,+} \otimes D^{k,+}$ or $D^{k,-} \otimes D^{k,-}$ [c.f. Fig.~\ref{fig:fixed-point_MPS}(b)]. If $k$ is odd, on the other hand, the congruence classes form the group $\mathbb{Z}_{4}$~\cite{duivenvoorden2013}, $[1, 1] + [1, 1] = [1, 3] + [1, 3] = [0, 2],~[1, 1] + [1, 3] = [1, 3] + [1, 1] = [0, 0]$, and the center is represented by $\rho_{\lambda} = e^{\frac{i\pi}{2}[\lambda]_{2}}\mathbb{I}$ that acts trivially on $D^{k,+} \otimes D^{k,-}$ or $D^{k,-} \otimes D^{k,+}$ [c.f. Fig.~\ref{fig:fixed-point_MPS}(c)]. Summarizing all the cases above, one comes to the conclusion that the center of $\mathrm{Spin}(n+1)$ always acts trivially on the local Hilbert spaces; for $n = 2k$ there are two distinct topological phases classified by $\mathbb{Z}_{2}$, whereas for $n = 2k - 1$ there are four phases which are classified by $\mathbb{Z}_{2} \times \mathbb{Z}_{2}$ (resp. $\mathbb{Z}_{4}$) if $k$ is even (resp. odd). These results can be understood physically in terms of the edge modes, as the latter transform in a projective representation of $G$ that belongs to a congruence class. This information is made manifest by the structure of the fixed-point MPSs (c.f. Fig.~\ref{fig:fixed-point_MPS}) since an arbitrary entanglement cut gives rise to a pair of virtual edge modes carrying the representation at an auxiliary site and the representation conjugate to it. These fixed-point MPSs are the representative states for the corresponding SPT phases, and those for all the other phases can be obtained by ``stacking'' these states (which amounts to taking tensor products of the representations).

\section{Kramers-Wannier duality and Onsager algebra}
\label{sec:duality}

The renowned Kramers-Wannier (KW) duality relates the physics of the same model at different couplings. The power of KW duality has been demonstrated in the context of two-dimensional Ising model (or TFI chain), where the transition temperature between ordered and disordered phases was identified as the self-dual point~\cite{kramers1941}. It is remarkable that this is an exact result which predates Onsager's exact solution~\cite{onsager1944}; the latter, in its original form, was based on an infinite-dimensional Lie algebra which now bears the name of Onsager and turns out to be closely related to the KW duality~\cite{dolan1982}. The aim of this Section is to show that our models given by~\eqref{eq:son-hamiltonian}, as natural generalizations of the TFI chain, also enjoy these elegant structures.

To this end, the Hamiltonian~\eqref{eq:son-hamiltonian} is rewritten as
\begin{equation}
    H = \sum_{j=1}^{N} \sum_{\alpha = 1}^{n} \sqrt{2} \left( h\mathbf{E}^{(\alpha)}_{2j-1} + J\mathbf{E}^{(\alpha)}_{2j} \right)
\end{equation}
up to a constant, with
\begin{equation}
    \mathbf{E}^{(\alpha)}_{2j-1} = \frac{1}{\sqrt{2}} \left( 1 - \Gamma^{2\alpha,2\alpha-1}_{j} \right), \quad j = 1, \ldots, N
\end{equation}
and
\begin{equation}
    \mathbf{E}^{(\alpha)}_{2j} = \begin{cases}
\begin{array}{c}
\frac{1}{\sqrt{2}} \left( 1 + \Gamma^{2\alpha-1,2n+1}_{j}\Gamma^{2\alpha}_{j+1} \right), \\
\frac{1}{\sqrt{2}} \left( 1 + \Gamma^{2\alpha-1,2n+1}_{N}\Gamma^{2\alpha}_{1} \right),
\end{array} & \begin{array}{c}
1 \leq j \leq N-1 \\
j = N.
\end{array} \end{cases}
\end{equation}
These operators generate $n$ independent copies of the (periodic) Temperley-Lieb algebra~\cite{temperley1971,levy1991}:
\begin{align}
\label{eq:TL-algebra}
    &(\mathbf{E}^{(\alpha)}_{j})^{2} = \sqrt{2} \mathbf{E}^{(\alpha)}_{j}, \quad \mathbf{E}^{(\alpha)}_{j} \mathbf{E}^{(\alpha)}_{j \pm 1~(\mathrm{mod}~2N)} \mathbf{E}^{(\alpha)}_{j} = \mathbf{E}^{(\alpha)}_{j}, \nonumber \\
    &\mathbf{E}^{(\alpha)}_{j} \mathbf{E}^{(\alpha)}_{j^{\prime}} = \mathbf{E}^{(\alpha)}_{j^{\prime}} \mathbf{E}^{(\alpha)}_{j}, \quad \vert j - j^{\prime}~(\mathrm{mod}~2N) \vert \geq 2,
\end{align}
where $\alpha = 1, \ldots, n$ and $j = 1, 2, \ldots, 2N$. For the case of TFI chain~\cite{montes2017}, the Temperley-Lieb generators are given by $\mathbf{E}_{2j-1} = ( 1 + \sigma^{x}_{j} )/\sqrt{2}$ and $\mathbf{E}_{2j} = ( 1 - \sigma^{z}_{j} \sigma^{z}_{j+1} )/\sqrt{2}$; the KW duality simply amounts to the map $\sigma^{x}_{j} \mapsto -\sigma^{z}_{j} \sigma^{z}_{j+1}$, $\sigma^{z}_{j} \sigma^{z}_{j+1} \mapsto -\sigma^{x}_{j+1}$ or $\mathbf{E}_{j} \mapsto \mathbf{E}_{j+1}$, with possible subtleties arising from the boundary condition. In fact, one can verify using~\eqref{eq:TL-algebra} that the unitary operator defined as
\begin{equation}
    U = \prod_{\alpha=1}^{n} \prod_{j=1}^{2N-1} \exp{\left( \frac{i\pi}{2\sqrt{2}}\mathbf{E}^{(\alpha)}_{j} \right)}
\end{equation}
acts on the Temperley-Lieb generators as
\begin{equation}
    U \mathbf{E}^{(\alpha)}_{j} U^{\dagger} = \begin{cases}
\begin{array}{c}
\mathbf{E}^{(\alpha)}_{j+1}, \quad 1 \leq j \leq 2N-2,\\
\frac{1}{\sqrt{2}} \left( 1 - w\Gamma^{2\alpha - 1, 2n + 1}_{N}\Gamma^{2\alpha}_{1} \right),~j = 2N-1,\\
\frac{1}{\sqrt{2}} \left( 1 + w\Gamma^{2\alpha,2\alpha-1}_{1} \right),~j = 2N.
\end{array} \end{cases}
\end{equation}
Thus, the unitary operator $U$ implements the duality transformation ($h \leftrightarrow J$), revealing directly on the lattice level the structure of exact KW duality in our models; the self-dual point at $J = h$ is precisely the $\mathrm{Spin}(n)_{1}$ quantum critical point. The only subtlety arises at the boundary; in particular, one finds that the form of the Hamiltonian is preserved by the duality transformation only in the sector with ``$\mathbb{Z}_{2}$ flux'' $w = -1$ (i.e., the NS sector). This is not surprising, as the KW duality is by no means any ``symmetry'' in the conventional sense. Nevertheless, it is worth noting that $[U,M^{\alpha\beta}] = 0$ with $M^{\alpha\beta} = \sum_{j=1}^{N} (\Gamma^{2\alpha-1,2\beta-1}_{j} + \Gamma^{2\alpha,2\beta}_{j}),~1 \leq \alpha < \beta \leq n$, so that the KW duality keeps the global $\mathrm{SO}(n)$ symmetry of our model intact.

The Onsager algebra was originally formulated in terms of an infinite number of generators satisfying certain commutation relations~\cite{onsager1944}. Equivalently, the Onsager algebra can be characterized by a pair of operators, $\mathbf{Q}$ and $\widetilde{\mathbf{Q}}$, subjecting to the so-called Dolan-Grady relations~\cite{dolan1982,davies1990,davies1991,perk2017}
\begin{equation}
\label{eq:dolan-grady}
    [\mathbf{Q}, [\mathbf{Q}, [\mathbf{Q}, \widetilde{\mathbf{Q}}]]] = 16 [\mathbf{Q}, \widetilde{\mathbf{Q}}],~[\widetilde{\mathbf{Q}}, [\widetilde{\mathbf{Q}}, [\widetilde{\mathbf{Q}}, \mathbf{Q}]]] = 16 [\widetilde{\mathbf{Q}}, \mathbf{Q}].
\end{equation}
By defining $\mathbf{Q} = \sum_{j=1}^{N} \sum_{\alpha = 1}^{n} \Gamma^{2\alpha-1, 2n+1}_{j} \Gamma^{2\alpha}_{j+1}$ and $\widetilde{\mathbf{Q}} = -\sum_{j=1}^{N} \sum_{\alpha = 1}^{n} \Gamma^{2\alpha, 2\alpha-1}_{j}$, the Hamiltonian~\eqref{eq:son-hamiltonian} simply reads $H = J\mathbf{Q} + h\widetilde{\mathbf{Q}}$ and it can be verified that $\mathbf{Q}$ and $\widetilde{\mathbf{Q}}$ fulfil the Dolan-Grady relations~\eqref{eq:dolan-grady}. Note that $\mathbf{Q}$ and $\widetilde{\mathbf{Q}}$ are related by the KW duality, under which~\eqref{eq:dolan-grady} is preserved. As a consequence of the Onsager algebra, an infinite set of conserved charges can be derived, providing insight into the integrability of our models.

\section{Summary and outlook}
\label{sec:summary}

In summary, we have constructed a family of SO($n$)-symmetric spin chains which generalize the transverse-field Ising chain for $n=1$. These models can be mapped to $n$ itinerant Majorana fermions coupled to a static $\mathbb{Z}_2$ gauge field and are hence exactly solvable. The phase diagram includes two distinct gapped phases as well as a critical point which is described by the $\mathrm{Spin}(n)_{1}$ CFT. One of the two gapped phases is trivial, while the other is an SPT phase. These two gapped phases are found to be related to each other via a Kramers-Wannier duality, while the $\mathrm{Spin}(n)_{1}$ critical point lies at the self-dual point. Closely related to the duality is the infinite-dimensional Onsager algebra; in fact, the interconnection among the Onsager algebra, the (generalized) Clifford algebra and the Temperley-Lieb algebra was exploited in Ref.~\cite{miao2022}. This reveals a rich algebraic structure of our models.

The nature of the quantum critical point was characterized by rigorously computing the partition function, of which the continuum limit agrees with the $\mathrm{Spin}(n)_{1}$ CFT. The latter CFT is formulated in terms of $n$ free massless Majorana fermion fields. In the vicinity of this critical point, the Majorana fermions acquire a non-zero but small mass; as the correlation length is relatively large, the continuous description still applies, for which the effective Hamiltonian density reads
\begin{equation}
\label{eq:hamiltonian-EFT}
    \mathcal{H}_{\textrm{eff}} = -i \sum_{\alpha=1}^{n} \left[ \frac{v}{2}\left( \xi^{\textrm{R}}_{\alpha}\partial_{x}\xi^{\textrm{R}}_{\alpha} - \xi^{\textrm{L}}_{\alpha}\partial_{x}\xi^{\textrm{L}}_{\alpha} \right) + m\xi^{\textrm{R}}_{\alpha}\xi^{\textrm{L}}_{\alpha} \right],
\end{equation}
where $\xi^{\textrm{R(L)}}_{\alpha}(x)$ is the right(left)-moving Majorana fermion field with ``color'' $\alpha~(= 1, \ldots, n)$, $v$ is the velocity, and $m$ is the Majorana mass. The masses of all Majorana fermion fields being the same implies SO($n$) symmetry.
The phase transition is indicated by the sign change of the Majorana mass. The models we proposed in this work furnish a perfect lattice realization of the effective field theory~\eqref{eq:hamiltonian-EFT}.

There are several interesting directions for future investigations. First, the SO(7) symmetry of the model with $n = 7$ can be explicitly broken down to $G_2$ by adding local interaction terms (see Ref.~\cite{li2022} for related discussions on a different model). It would be worth exploring how to use our model as the basic building block for constructing the Fibonacci topological superconductor proposed by Hu and Kane~\cite{hu2018}. Second, we expect that the method for constructing exactly solvable models with Majorana fermions coupled to $\mathbb{Z}_2$ gauge fields can also be used for proposing symmetry-protected quantum critical models~\cite{verresen2018,parker2018,jones2019,verresen2021,jones2022}. It would be interesting to explore, for instance, duality~\cite{yang2023,li2023} and conformal boundary conditions~\cite{yu2022} in such models. Third, one may consider multiple species of $\mathbb{Z}_n$ parafermions coupled to a static $\mathbb{Z}_n$ gauge field, which would be a natural generalization of the models in the present work. It is expected that exotic critical points and different types of gapped phases would emerge. Finally, the beautiful formalism of topological defects~\cite{aasen2016,aasen2020,lootens2021} may also be exploited to shed light on the Kramers-Wannier duality or, more generally, categorical symmetries in our models; this aspect will be considered in future works.

\emph{Note added.}- After submitting the preprint of our manuscript to arXiv, we were reminded by the authors of Ref.~\cite{chugh2022} that they have also considered one-dimensional Gamma matrix models including our Hamiltonian~\eqref{eq:son-hamiltonian} and solved them using the Jordan-Wigner transformation. This calls for a comparison between our present work and Ref.~\cite{chugh2022}. The analysis of the conformal criticalities and the nature of the gapped phases, which constitutes an essential part of our present work, was not included in Ref.~\cite{chugh2022}. Moreover, both works complement each other as the method being used in our work to solve the models is Kitaev-type Majorana representation, in contrast to the Jordan-Wigner transformation. We thank the authors of Ref.~\cite{chugh2022} for bringing their work to our attention.

\acknowledgements

We are grateful to Xiao-Yu Dong, Lukas Janssen, Urban Seifert, and Matthias Vojta for stimulating discussions and collaborations on related topics. We also thank Meng Cheng and Yuan Miao for helpful discussions. This research has been funded by the Deutsche Forschungsgemeinschaft (DFG) through project
A06 of SFB 1143 (project No.~247310070) and the IMPRS for Quantum Dynamics and Control at MPI-PKS. H.C.Z. acknowledges a scholarship from the CAS-DAAD Doctoral Joint Fellowship Program.
\appendix

\section{Representation of the Clifford algebra $\mathrm{Cl}_{2n+1,0}(\mathbb{R})$}
\label{append:clifford-rep}

To construct a representation of the Clifford algebra $\mathrm{Cl}_{2n+1,0}(\mathbb{R})$, we distinguish between the cases $n = 2k$ and $n = 2k - 1$. To make the connection with spinor representations of the $\mathrm{so}(n+1)$ algebra clear (as exploited in Sec.~\ref{sec:phases}), in the following we utilize two copies of representation of the Clifford algebra $\mathrm{Cl}_{2k+1,0}(\mathbb{R})$ to construct that of $\mathrm{Cl}_{2n+1,0}(\mathbb{R})$.

It is straightforward to verify that the $2^{k}$-dimensional matrices $\Lambda^{\alpha},~\alpha = 1, 2, \ldots, (2k+1),$ defined as~\cite{georgi1999}
\begin{widetext}
\begin{align}
\label{eq:lambda-definition}
    \Lambda^{1} =&~ \sigma^{y}_{1} \otimes \sigma^{z}_{2} \otimes \cdots \otimes \sigma^{z}_{k}, \qquad \Lambda^{2} = -\sigma^{x}_{1} \otimes \sigma^{z}_{2} \otimes \cdots \otimes \sigma^{z}_{k}, \nonumber \\
    \Lambda^{3} =&~ \sigma^{0}_{1} \otimes \sigma^{y}_{2} \otimes \cdots \otimes \sigma^{z}_{k}, \qquad \Lambda^{4} = -\sigma^{0}_{1} \otimes \sigma^{x}_{2} \otimes \cdots \otimes \sigma^{z}_{k}, \nonumber \\
    &\cdots~\cdots \nonumber \\
    \Lambda^{2k-1} =&~ \sigma^{0}_{1} \otimes \sigma^{0}_{2} \otimes \cdots \otimes \sigma^{y}_{k}, \qquad \Lambda^{2k} = -\sigma^{0}_{1} \otimes \sigma^{0}_{2} \otimes \cdots \otimes \sigma^{x}_{k}, \nonumber \\
    \Lambda^{2k+1} =&~ \sigma^{z}_{1} \otimes \sigma^{z}_{2} \otimes \cdots \otimes \sigma^{z}_{k},
\end{align}
\end{widetext}
satisfy $\{ \Lambda^{\alpha}, \Lambda^{\beta} \} = 2\delta_{\alpha\beta}$ and generate $\mathrm{Cl}_{2k+1,0}(\mathbb{R})$. Here $\sigma^{0}$ is the $2 \times 2$ identity matrix and the subscripts label the Hilbert subspaces in which these matrices act. It is also convenient to define the commutators $\Lambda^{\alpha\beta} \equiv \frac{i}{2} [\Lambda^{\alpha},\Lambda^{\beta}]$ among these matrices. As we will be using two copies of this representation of $\mathrm{Cl}_{2k+1,0}(\mathbb{R})$, let us denote the matrices defined in~\eqref{eq:lambda-definition} as $\Lambda^{\alpha}_{1 \rightarrow k}$, and $\Lambda^{\alpha}_{k+1 \rightarrow 2k}$ are similarly defined.

\subsection{$n = 2k$}

For the case $n = 2k$, the $(2n + 1)$ Gamma-matrices given by
\begin{align}
\label{eq:def-gamma-2k}
    \Gamma^{2\alpha-1} &= \Lambda^{2k+1}_{1 \rightarrow k} \otimes \Lambda^{\alpha}_{k+1 \rightarrow 2k}, \quad \alpha = 1, 2, \ldots, (2k+1), \nonumber \\
    \Gamma^{2\alpha} &= \Lambda^{\alpha, 2k+1}_{1 \rightarrow k} \otimes \boldsymbol{1}_{k+1 \rightarrow 2k}, \quad \alpha = 1, 2, \ldots, 2k
\end{align}
generate $\mathrm{Cl}_{2n+1,0}(\mathbb{R})$, where $\boldsymbol{1}_{k+1 \rightarrow 2k} \equiv \sigma^{0}_{k+1} \otimes \cdots \otimes \sigma^{0}_{2k}$. These Gamma-matrices are $2^{2k}$-dimensional; according to the definition~\eqref{eq:def-gamma-2k}, their product can be computed as follows:
\begin{align}
    &\Gamma^{1}\Gamma^{2}\cdots\Gamma^{2n+1} \nonumber \\
    =&~(-1)^{k} \left( \prod_{\alpha=1}^{2k} \Gamma^{2\alpha-1} \right) \left( \prod_{\alpha=1}^{2k} \Gamma^{2\alpha} \right) \Gamma^{4k+1} \nonumber \\
    =&~(-1)^{k} \left( \prod_{\alpha=1}^{2k+1} \Lambda^{\alpha}_{1 \rightarrow k} \right) \otimes \left( \prod_{\alpha=1}^{2k+1} \Lambda^{\alpha}_{k+1 \rightarrow 2k} \right) \nonumber \\
    =&~\boldsymbol{1}_{1 \rightarrow 2k},
\end{align}
where we have made use of the identity $\Lambda^{1}\Lambda^{2}\cdots\Lambda^{2k+1} = i^{k}$ in the last step. Thus, the local fermion parity~\eqref{eq:local-fermion-parity} is $Q_{\textrm{loc.}} = (-1)^{k}$ within this representation. Using~\eqref{eq:def-gamma-2k}, the representation for other operators that appear in the Hamiltonian~\eqref{eq:son-hamiltonian} can also be derived:
\begin{equation}
    \Gamma^{2\alpha-1,2n+1} = i \Gamma^{2\alpha-1} \Gamma^{2n+1} = \boldsymbol{1}_{1 \rightarrow k} \otimes \Lambda^{\alpha,2k+1}_{k+1 \rightarrow 2k},
\end{equation}
\begin{equation}
    \Gamma^{2\alpha,2\alpha-1} = i \Gamma^{2\alpha} \Gamma^{2\alpha-1} = - \Lambda^{\alpha}_{1 \rightarrow k} \otimes \Lambda^{\alpha}_{k+1 \rightarrow 2k},
\end{equation}
where $\alpha = 1, 2, \ldots, 2k$.

To illustrate the above representation more concretely, let us consider the simplest example with $k=1$ and $n=2$. In this case, the Gamma-matrices are expressed in terms of the Pauli operators as follows:
\begin{align}
    &\Gamma^{1} = \sigma^{z} \otimes \sigma^{y}, \quad \Gamma^{2} = -\sigma^{x} \otimes \sigma^{0}, \nonumber \\
    &\Gamma^{3} = -\sigma^{z} \otimes \sigma^{x}, \quad \Gamma^{4} = -\sigma^{y} \otimes \sigma^{0}, \nonumber \\
    &\Gamma^{5} = \sigma^{z} \otimes \sigma^{z}.
\end{align}
To simplify the Hamiltonian~\eqref{eq:son-hamiltonian} with $n=2$, it is useful to regard each site in the chain as composed of two ``constituent'' ones; the four-dimensional local Hilbert space at each original site is accordingly decomposed into the tensor product of two two-dimensional ones. After this ``splitting'' transformation, the Hamiltonian reads
\begin{align}
    H_{n=2} = &\sum_{j=1}^{N} \Big[ J \left( \sigma^{x}_{2j} \sigma^{x}_{2j+1} + \sigma^{y}_{2j} \sigma^{y}_{2j+1} \right) \nonumber \\
    &+ h \left( \sigma^{x}_{2j-1} \sigma^{x}_{2j} + \sigma^{y}_{2j-1} \sigma^{y}_{2j} \right) \Big],
\end{align}
which is that of a spin-$1/2$ bond-alternating isotropic XY chain, which further reduces to the homogeneous one at the critical point $J = h$.

\subsection{$n = 2k - 1$}

In this case, the Gamma-matrices are represented in parallel with the case $n = 2k$:
\begin{align}
    \Gamma^{2\alpha-1} &= \Lambda^{2k}_{1 \rightarrow k} \otimes \Lambda^{\alpha}_{k+1 \rightarrow 2k}, \quad \alpha = 1, 2, \ldots, 2k, \nonumber \\
    \Gamma^{2\alpha} &= \Lambda^{\alpha, 2k}_{1 \rightarrow k} \otimes \boldsymbol{1}_{k+1 \rightarrow 2k}, \quad \alpha = 1, 2, \ldots, (2k-1).
\end{align}
The product of these Gamma-matrices are now
\begin{equation}
    \Gamma^{1}\Gamma^{2}\cdots\Gamma^{2n+1} = i \sigma^{z}_{1} \otimes \sigma^{z}_{2} \otimes \cdots \otimes \sigma^{z}_{2k}.
\end{equation}
Apparently, $Q_{\textrm{loc.}} = i^{2k-1} \Gamma^{1}\Gamma^{2}\cdots\Gamma^{2n+1}$ is diagonal and $Q_{\textrm{loc.}}^2 = 1$. The representation for other operators is similarly derived:
\begin{equation}
    \Gamma^{2\alpha-1,2n+1} = \boldsymbol{1}_{1 \rightarrow k} \otimes \Lambda^{\alpha,2k}_{k+1 \rightarrow 2k},
\end{equation}
\begin{equation}
    \Gamma^{2\alpha,2\alpha-1} = - \Lambda^{\alpha}_{1 \rightarrow k} \otimes \Lambda^{\alpha}_{k+1 \rightarrow 2k},
\end{equation}
where $\alpha = 1, 2, \ldots, (2k-1)$.

Note that the dimension of the above Gamma-matrices is $2^{2k}$, which is the same as that in the case $n = 2k$. However, as $Q_{\textrm{loc.}}$ commutes with the set $\{ \Gamma^{1}, \Gamma^{2}, \ldots, \Gamma^{2n+1} \}$, each of the Gamma-matrices breaks up into two blocks of size $2^{2k-1} \times 2^{2k-1}$ ($= 2^{n} \times 2^{n}$) that are associated with the local fermion parity eigenvalues $Q_{\textrm{loc.}} = \pm 1$, respectively. Let us now illustrate this point by considering the simplest example with $k = 1$ and $n = 1$. In this case, one has (zero entries are left empty)
\begin{align}
    &\Gamma^{1} = -\sigma^{x} \otimes \sigma^{y} = \left( \begin{array}{cccc}
        ~& ~& ~& i \\
        ~& ~& -i& ~ \\
        ~& i& ~& ~ \\
        -i& ~& ~& ~
    \end{array} \right), \nonumber \\
    &\Gamma^{2} = -\sigma^{z} \otimes \sigma^{0} = \mathrm{diag}(-1, -1,~1,~1), \nonumber \\
    &\Gamma^{3} = \sigma^{x} \otimes \sigma^{x} = \left( \begin{array}{cccc}
        ~& ~& ~& 1 \\
        ~& ~& 1& ~ \\
        ~& 1& ~& ~ \\
        1& ~& ~& ~
    \end{array} \right),
\end{align}
and $Q_{\textrm{loc.}} = -\sigma^{z} \otimes \sigma^{z} = \mathrm{diag}(-1,~1,~1,-1)$. By projecting into the subspace with $Q_{\textrm{loc.}} = +1$ [which amounts to applying with the projector $(1 + Q_{\textrm{loc.}})/2$], one finds that $\Gamma^{1} \mapsto \sigma^{y}$, $\Gamma^{2} \mapsto -\sigma^{z}$, $\Gamma^{3} \mapsto \sigma^{x}$, and other operators that appear in the Hamiltonian~\eqref{eq:son-hamiltonian} with $n = 1$ become $\Gamma^{1,3} \mapsto \sigma^{z},~\Gamma^{2,1} \mapsto -\sigma^{x}$. Thus, we conclude that $H_{n=1}$ is nothing but the Hamiltonian of the TFI chain given in~\eqref{eq:TFIM-hamiltonian}.

\bibliography{chain.bib}

\begin{thebibliography}{86}%
\makeatletter
\providecommand \@ifxundefined [1]{%
 \@ifx{#1\undefined}
}%
\providecommand \@ifnum [1]{%
 \ifnum #1\expandafter \@firstoftwo
 \else \expandafter \@secondoftwo
 \fi
}%
\providecommand \@ifx [1]{%
 \ifx #1\expandafter \@firstoftwo
 \else \expandafter \@secondoftwo
 \fi
}%
\providecommand \natexlab [1]{#1}%
\providecommand \enquote  [1]{``#1''}%
\providecommand \bibnamefont  [1]{#1}%
\providecommand \bibfnamefont [1]{#1}%
\providecommand \citenamefont [1]{#1}%
\providecommand \href@noop [0]{\@secondoftwo}%
\providecommand \href [0]{\begingroup \@sanitize@url \@href}%
\providecommand \@href[1]{\@@startlink{#1}\@@href}%
\providecommand \@@href[1]{\endgroup#1\@@endlink}%
\providecommand \@sanitize@url [0]{\catcode `\\12\catcode `\$12\catcode
  `\&12\catcode `\#12\catcode `\^12\catcode `\_12\catcode `\%12\relax}%
\providecommand \@@startlink[1]{}%
\providecommand \@@endlink[0]{}%
\providecommand \url  [0]{\begingroup\@sanitize@url \@url }%
\providecommand \@url [1]{\endgroup\@href {#1}{\urlprefix }}%
\providecommand \urlprefix  [0]{URL }%
\providecommand \Eprint [0]{\href }%
\providecommand \doibase [0]{https://doi.org/}%
\providecommand \selectlanguage [0]{\@gobble}%
\providecommand \bibinfo  [0]{\@secondoftwo}%
\providecommand \bibfield  [0]{\@secondoftwo}%
\providecommand \translation [1]{[#1]}%
\providecommand \BibitemOpen [0]{}%
\providecommand \bibitemStop [0]{}%
\providecommand \bibitemNoStop [0]{.\EOS\space}%
\providecommand \EOS [0]{\spacefactor3000\relax}%
\providecommand \BibitemShut  [1]{\csname bibitem#1\endcsname}%
\let\auto@bib@innerbib\@empty
\bibitem [{\citenamefont {Cardy}(1996)}]{cardy1996}%
  \BibitemOpen
  \bibfield  {author} {\bibinfo {author} {\bibfnamefont {J.}~\bibnamefont
  {Cardy}},\ }\href@noop {} {\emph {\bibinfo {title} {Scaling and
  Renormalization in Statistical Physics}}}\ (\bibinfo  {publisher} {Cambridge
  University Press, Cambridge},\ \bibinfo {year} {1996})\BibitemShut {NoStop}%
\bibitem [{\citenamefont {Sachdev}(2011)}]{sachdev2011}%
  \BibitemOpen
  \bibfield  {author} {\bibinfo {author} {\bibfnamefont {S.}~\bibnamefont
  {Sachdev}},\ }\href@noop {} {\emph {\bibinfo {title} {Quantum Phase
  Transitions}}}\ (\bibinfo  {publisher} {Cambridge University Press,
  Cambridge},\ \bibinfo {year} {2011})\BibitemShut {NoStop}%
\bibitem [{\citenamefont {Mussardo}(2020)}]{mussardo2020}%
  \BibitemOpen
  \bibfield  {author} {\bibinfo {author} {\bibfnamefont {G.}~\bibnamefont
  {Mussardo}},\ }\href@noop {} {\emph {\bibinfo {title} {Statistical Field
  Theory: An Introduction to Exactly Solved Models in Statistical Physics}}}\
  (\bibinfo  {publisher} {Oxford University Press, Oxford},\ \bibinfo {year}
  {2020})\BibitemShut {NoStop}%
\bibitem [{\citenamefont {Kramers}\ and\ \citenamefont
  {Wannier}(1941)}]{kramers1941}%
  \BibitemOpen
  \bibfield  {author} {\bibinfo {author} {\bibfnamefont {H.~A.}\ \bibnamefont
  {Kramers}}\ and\ \bibinfo {author} {\bibfnamefont {G.~H.}\ \bibnamefont
  {Wannier}},\ }\href {https://doi.org/10.1103/PhysRev.60.252} {\bibfield
  {journal} {\bibinfo  {journal} {Phys. Rev.}\ }\textbf {\bibinfo {volume}
  {60}},\ \bibinfo {pages} {252} (\bibinfo {year} {1941})}\BibitemShut
  {NoStop}%
\bibitem [{\citenamefont {El-Chaar}()}]{el-chaar2012}%
  \BibitemOpen
  \bibfield  {author} {\bibinfo {author} {\bibfnamefont {C.}~\bibnamefont
  {El-Chaar}},\ }\href {https://arxiv.org/abs/1205.5989} {\ }\Eprint
  {https://arxiv.org/abs/1205.5989} {arXiv:1205.5989} \BibitemShut {NoStop}%
\bibitem [{\citenamefont {Onsager}(1944)}]{onsager1944}%
  \BibitemOpen
  \bibfield  {author} {\bibinfo {author} {\bibfnamefont {L.}~\bibnamefont
  {Onsager}},\ }\href {https://doi.org/10.1103/PhysRev.65.117} {\bibfield
  {journal} {\bibinfo  {journal} {Phys. Rev.}\ }\textbf {\bibinfo {volume}
  {65}},\ \bibinfo {pages} {117} (\bibinfo {year} {1944})}\BibitemShut
  {NoStop}%
\bibitem [{\citenamefont {Fradkin}\ and\ \citenamefont
  {Susskind}(1978)}]{fradkin1978}%
  \BibitemOpen
  \bibfield  {author} {\bibinfo {author} {\bibfnamefont {E.}~\bibnamefont
  {Fradkin}}\ and\ \bibinfo {author} {\bibfnamefont {L.}~\bibnamefont
  {Susskind}},\ }\href {https://doi.org/10.1103/PhysRevD.17.2637} {\bibfield
  {journal} {\bibinfo  {journal} {Phys. Rev. D}\ }\textbf {\bibinfo {volume}
  {17}},\ \bibinfo {pages} {2637} (\bibinfo {year} {1978})}\BibitemShut
  {NoStop}%
\bibitem [{\citenamefont {Kogut}(1979)}]{kogut1979}%
  \BibitemOpen
  \bibfield  {author} {\bibinfo {author} {\bibfnamefont {J.~B.}\ \bibnamefont
  {Kogut}},\ }\href {https://doi.org/10.1103/RevModPhys.51.659} {\bibfield
  {journal} {\bibinfo  {journal} {Rev. Mod. Phys.}\ }\textbf {\bibinfo {volume}
  {51}},\ \bibinfo {pages} {659} (\bibinfo {year} {1979})}\BibitemShut
  {NoStop}%
\bibitem [{\citenamefont {Dolan}\ and\ \citenamefont
  {Grady}(1982)}]{dolan1982}%
  \BibitemOpen
  \bibfield  {author} {\bibinfo {author} {\bibfnamefont {L.}~\bibnamefont
  {Dolan}}\ and\ \bibinfo {author} {\bibfnamefont {M.}~\bibnamefont {Grady}},\
  }\href {https://doi.org/10.1103/PhysRevD.25.1587} {\bibfield  {journal}
  {\bibinfo  {journal} {Phys. Rev. D}\ }\textbf {\bibinfo {volume} {25}},\
  \bibinfo {pages} {1587} (\bibinfo {year} {1982})}\BibitemShut {NoStop}%
\bibitem [{\citenamefont {Ortiz}\ \emph {et~al.}(2012)\citenamefont {Ortiz},
  \citenamefont {Cobanera},\ and\ \citenamefont {Nussinov}}]{ortiz2012}%
  \BibitemOpen
  \bibfield  {author} {\bibinfo {author} {\bibfnamefont {G.}~\bibnamefont
  {Ortiz}}, \bibinfo {author} {\bibfnamefont {E.}~\bibnamefont {Cobanera}},\
  and\ \bibinfo {author} {\bibfnamefont {Z.}~\bibnamefont {Nussinov}},\ }\href
  {https://doi.org/10.1016/j.nuclphysb.2011.09.012} {\bibfield  {journal}
  {\bibinfo  {journal} {Nucl. Phys. B}\ }\textbf {\bibinfo {volume} {854}},\
  \bibinfo {pages} {780} (\bibinfo {year} {2012})}\BibitemShut {NoStop}%
\bibitem [{\citenamefont {Takhtajan}(1982)}]{takhtajan1982}%
  \BibitemOpen
  \bibfield  {author} {\bibinfo {author} {\bibfnamefont {L.}~\bibnamefont
  {Takhtajan}},\ }\href {https://doi.org/10.1016/0375-9601(82)90764-2}
  {\bibfield  {journal} {\bibinfo  {journal} {Phys. Letts. A}\ }\textbf
  {\bibinfo {volume} {87}},\ \bibinfo {pages} {479} (\bibinfo {year}
  {1982})}\BibitemShut {NoStop}%
\bibitem [{\citenamefont {Babujian}(1982)}]{babujian1982}%
  \BibitemOpen
  \bibfield  {author} {\bibinfo {author} {\bibfnamefont {H.}~\bibnamefont
  {Babujian}},\ }\href {https://doi.org/10.1016/0375-9601(82)90403-0}
  {\bibfield  {journal} {\bibinfo  {journal} {Phys. Letts. A}\ }\textbf
  {\bibinfo {volume} {90}},\ \bibinfo {pages} {479} (\bibinfo {year}
  {1982})}\BibitemShut {NoStop}%
\bibitem [{\citenamefont {Affleck}\ and\ \citenamefont
  {Haldane}(1987)}]{affleck1987}%
  \BibitemOpen
  \bibfield  {author} {\bibinfo {author} {\bibfnamefont {I.}~\bibnamefont
  {Affleck}}\ and\ \bibinfo {author} {\bibfnamefont {F.~D.~M.}\ \bibnamefont
  {Haldane}},\ }\href {https://doi.org/10.1103/PhysRevB.36.5291} {\bibfield
  {journal} {\bibinfo  {journal} {Phys. Rev. B}\ }\textbf {\bibinfo {volume}
  {36}},\ \bibinfo {pages} {5291} (\bibinfo {year} {1987})}\BibitemShut
  {NoStop}%
\bibitem [{\citenamefont {Tsvelik}(1990)}]{tsvelik1990}%
  \BibitemOpen
  \bibfield  {author} {\bibinfo {author} {\bibfnamefont {A.~M.}\ \bibnamefont
  {Tsvelik}},\ }\href {https://doi.org/10.1103/PhysRevB.42.10499} {\bibfield
  {journal} {\bibinfo  {journal} {Phys. Rev. B}\ }\textbf {\bibinfo {volume}
  {42}},\ \bibinfo {pages} {10499} (\bibinfo {year} {1990})}\BibitemShut
  {NoStop}%
\bibitem [{\citenamefont {Lecheminant}\ and\ \citenamefont
  {Orignac}(2002)}]{lecheminant2002}%
  \BibitemOpen
  \bibfield  {author} {\bibinfo {author} {\bibfnamefont {P.}~\bibnamefont
  {Lecheminant}}\ and\ \bibinfo {author} {\bibfnamefont {E.}~\bibnamefont
  {Orignac}},\ }\href {https://doi.org/10.1103/PhysRevB.65.174406} {\bibfield
  {journal} {\bibinfo  {journal} {Phys. Rev. B}\ }\textbf {\bibinfo {volume}
  {65}},\ \bibinfo {pages} {174406} (\bibinfo {year} {2002})}\BibitemShut
  {NoStop}%
\bibitem [{\citenamefont {Liu}\ \emph {et~al.}(2012)\citenamefont {Liu},
  \citenamefont {Zhou}, \citenamefont {Tu}, \citenamefont {Wen},\ and\
  \citenamefont {Ng}}]{liu2012}%
  \BibitemOpen
  \bibfield  {author} {\bibinfo {author} {\bibfnamefont {Z.-X.}\ \bibnamefont
  {Liu}}, \bibinfo {author} {\bibfnamefont {Y.}~\bibnamefont {Zhou}}, \bibinfo
  {author} {\bibfnamefont {H.-H.}\ \bibnamefont {Tu}}, \bibinfo {author}
  {\bibfnamefont {X.-G.}\ \bibnamefont {Wen}},\ and\ \bibinfo {author}
  {\bibfnamefont {T.-K.}\ \bibnamefont {Ng}},\ }\href
  {https://doi.org/10.1103/PhysRevB.85.195144} {\bibfield  {journal} {\bibinfo
  {journal} {Phys. Rev. B}\ }\textbf {\bibinfo {volume} {85}},\ \bibinfo
  {pages} {195144} (\bibinfo {year} {2012})}\BibitemShut {NoStop}%
\bibitem [{\citenamefont {Liu}\ \emph {et~al.}(2014)\citenamefont {Liu},
  \citenamefont {Zhou},\ and\ \citenamefont {Ng}}]{liu2014}%
  \BibitemOpen
  \bibfield  {author} {\bibinfo {author} {\bibfnamefont {Z.-X.}\ \bibnamefont
  {Liu}}, \bibinfo {author} {\bibfnamefont {Y.}~\bibnamefont {Zhou}},\ and\
  \bibinfo {author} {\bibfnamefont {T.-K.}\ \bibnamefont {Ng}},\ }\href
  {https://doi.org/10.1088/1367-2630/16/8/083031} {\bibfield  {journal}
  {\bibinfo  {journal} {New J. Phys.}\ }\textbf {\bibinfo {volume} {16}},\
  \bibinfo {pages} {083031} (\bibinfo {year} {2014})}\BibitemShut {NoStop}%
\bibitem [{\citenamefont {Chen}()}]{chen2015}%
  \BibitemOpen
  \bibfield  {author} {\bibinfo {author} {\bibfnamefont {K.-T.}\ \bibnamefont
  {Chen}},\ }\href {https://arxiv.org/abs/1512.00967} {\ }\Eprint
  {https://arxiv.org/abs/1512.00967} {arXiv:1512.00967} \BibitemShut {NoStop}%
\bibitem [{\citenamefont {Shelton}\ \emph {et~al.}(1996)\citenamefont
  {Shelton}, \citenamefont {Nersesyan},\ and\ \citenamefont
  {Tsvelik}}]{shelton1996}%
  \BibitemOpen
  \bibfield  {author} {\bibinfo {author} {\bibfnamefont {D.~G.}\ \bibnamefont
  {Shelton}}, \bibinfo {author} {\bibfnamefont {A.~A.}\ \bibnamefont
  {Nersesyan}},\ and\ \bibinfo {author} {\bibfnamefont {A.~M.}\ \bibnamefont
  {Tsvelik}},\ }\href {https://doi.org/10.1103/PhysRevB.53.8521} {\bibfield
  {journal} {\bibinfo  {journal} {Phys. Rev. B}\ }\textbf {\bibinfo {volume}
  {53}},\ \bibinfo {pages} {8521} (\bibinfo {year} {1996})}\BibitemShut
  {NoStop}%
\bibitem [{\citenamefont {Nersesyan}\ and\ \citenamefont
  {Tsvelik}(1997)}]{nersesyan1997}%
  \BibitemOpen
  \bibfield  {author} {\bibinfo {author} {\bibfnamefont {A.~A.}\ \bibnamefont
  {Nersesyan}}\ and\ \bibinfo {author} {\bibfnamefont {A.~M.}\ \bibnamefont
  {Tsvelik}},\ }\href {https://doi.org/10.1103/PhysRevLett.78.3939} {\bibfield
  {journal} {\bibinfo  {journal} {Phys. Rev. Lett.}\ }\textbf {\bibinfo
  {volume} {78}},\ \bibinfo {pages} {3939} (\bibinfo {year}
  {1997})}\BibitemShut {NoStop}%
\bibitem [{\citenamefont {Tu}\ \emph {et~al.}(2008{\natexlab{a}})\citenamefont
  {Tu}, \citenamefont {Zhang},\ and\ \citenamefont {Xiang}}]{tu2008a}%
  \BibitemOpen
  \bibfield  {author} {\bibinfo {author} {\bibfnamefont {H.-H.}\ \bibnamefont
  {Tu}}, \bibinfo {author} {\bibfnamefont {G.-M.}\ \bibnamefont {Zhang}},\ and\
  \bibinfo {author} {\bibfnamefont {T.}~\bibnamefont {Xiang}},\ }\href
  {https://doi.org/10.1103/PhysRevB.78.094404} {\bibfield  {journal} {\bibinfo
  {journal} {Phys. Rev. B}\ }\textbf {\bibinfo {volume} {78}},\ \bibinfo
  {pages} {094404} (\bibinfo {year} {2008}{\natexlab{a}})}\BibitemShut
  {NoStop}%
\bibitem [{\citenamefont {Tu}\ \emph {et~al.}(2008{\natexlab{b}})\citenamefont
  {Tu}, \citenamefont {Zhang},\ and\ \citenamefont {Xiang}}]{tu2008b}%
  \BibitemOpen
  \bibfield  {author} {\bibinfo {author} {\bibfnamefont {H.-H.}\ \bibnamefont
  {Tu}}, \bibinfo {author} {\bibfnamefont {G.-M.}\ \bibnamefont {Zhang}},\ and\
  \bibinfo {author} {\bibfnamefont {T.}~\bibnamefont {Xiang}},\ }\href
  {https://doi.org/10.1088/1751-8113/41/41/415201} {\bibfield  {journal}
  {\bibinfo  {journal} {J. Phys. A: Math. Theor.}\ }\textbf {\bibinfo {volume}
  {41}},\ \bibinfo {pages} {415201} (\bibinfo {year}
  {2008}{\natexlab{b}})}\BibitemShut {NoStop}%
\bibitem [{\citenamefont {Tu}\ and\ \citenamefont {Or\'us}(2011)}]{tu2011}%
  \BibitemOpen
  \bibfield  {author} {\bibinfo {author} {\bibfnamefont {H.-H.}\ \bibnamefont
  {Tu}}\ and\ \bibinfo {author} {\bibfnamefont {R.}~\bibnamefont {Or\'us}},\
  }\href {https://doi.org/10.1103/PhysRevLett.107.077204} {\bibfield  {journal}
  {\bibinfo  {journal} {Phys. Rev. Lett.}\ }\textbf {\bibinfo {volume} {107}},\
  \bibinfo {pages} {077204} (\bibinfo {year} {2011})}\BibitemShut {NoStop}%
\bibitem [{\citenamefont {Alet}\ \emph {et~al.}(2011)\citenamefont {Alet},
  \citenamefont {Capponi}, \citenamefont {Nonne}, \citenamefont {Lecheminant},\
  and\ \citenamefont {McCulloch}}]{alet2011}%
  \BibitemOpen
  \bibfield  {author} {\bibinfo {author} {\bibfnamefont {F.}~\bibnamefont
  {Alet}}, \bibinfo {author} {\bibfnamefont {S.}~\bibnamefont {Capponi}},
  \bibinfo {author} {\bibfnamefont {H.}~\bibnamefont {Nonne}}, \bibinfo
  {author} {\bibfnamefont {P.}~\bibnamefont {Lecheminant}},\ and\ \bibinfo
  {author} {\bibfnamefont {I.~P.}\ \bibnamefont {McCulloch}},\ }\href
  {https://doi.org/10.1103/PhysRevB.83.060407} {\bibfield  {journal} {\bibinfo
  {journal} {Phys. Rev. B}\ }\textbf {\bibinfo {volume} {83}},\ \bibinfo
  {pages} {060407(R)} (\bibinfo {year} {2011})}\BibitemShut {NoStop}%
\bibitem [{\citenamefont {Okunishi}\ and\ \citenamefont
  {Harada}(2014)}]{okunishi2014}%
  \BibitemOpen
  \bibfield  {author} {\bibinfo {author} {\bibfnamefont {K.}~\bibnamefont
  {Okunishi}}\ and\ \bibinfo {author} {\bibfnamefont {K.}~\bibnamefont
  {Harada}},\ }\href {https://doi.org/10.1103/PhysRevB.89.134422} {\bibfield
  {journal} {\bibinfo  {journal} {Phys. Rev. B}\ }\textbf {\bibinfo {volume}
  {89}},\ \bibinfo {pages} {134422} (\bibinfo {year} {2014})}\BibitemShut
  {NoStop}%
\bibitem [{\citenamefont {Kitaev}(2006)}]{kitaev2006}%
  \BibitemOpen
  \bibfield  {author} {\bibinfo {author} {\bibfnamefont {A.}~\bibnamefont
  {Kitaev}},\ }\href {https://doi.org/10.1016/j.aop.2005.10.005} {\bibfield
  {journal} {\bibinfo  {journal} {Ann. Phys.}\ }\textbf {\bibinfo {volume}
  {321}},\ \bibinfo {pages} {2 } (\bibinfo {year} {2006})}\BibitemShut
  {NoStop}%
\bibitem [{\citenamefont {Chulliparambil}\ \emph {et~al.}(2020)\citenamefont
  {Chulliparambil}, \citenamefont {Seifert}, \citenamefont {Vojta},
  \citenamefont {Janssen},\ and\ \citenamefont {Tu}}]{chulliparambil2020}%
  \BibitemOpen
  \bibfield  {author} {\bibinfo {author} {\bibfnamefont {S.}~\bibnamefont
  {Chulliparambil}}, \bibinfo {author} {\bibfnamefont {U.~F.~P.}\ \bibnamefont
  {Seifert}}, \bibinfo {author} {\bibfnamefont {M.}~\bibnamefont {Vojta}},
  \bibinfo {author} {\bibfnamefont {L.}~\bibnamefont {Janssen}},\ and\ \bibinfo
  {author} {\bibfnamefont {H.-H.}\ \bibnamefont {Tu}},\ }\href
  {https://doi.org/10.1103/PhysRevB.102.201111} {\bibfield  {journal} {\bibinfo
   {journal} {Phys. Rev. B}\ }\textbf {\bibinfo {volume} {102}},\ \bibinfo
  {pages} {201111(R)} (\bibinfo {year} {2020})}\BibitemShut {NoStop}%
\bibitem [{\citenamefont {Pfeuty}(1970)}]{pfeuty1970}%
  \BibitemOpen
  \bibfield  {author} {\bibinfo {author} {\bibfnamefont {P.}~\bibnamefont
  {Pfeuty}},\ }\href {https://doi.org/10.1016/0003-4916(70)90270-8} {\bibfield
  {journal} {\bibinfo  {journal} {Ann. Phys.}\ }\textbf {\bibinfo {volume}
  {57}},\ \bibinfo {pages} {79} (\bibinfo {year} {1970})}\BibitemShut {NoStop}%
\bibitem [{\citenamefont {Kitaev}(2001)}]{kitaev2001}%
  \BibitemOpen
  \bibfield  {author} {\bibinfo {author} {\bibfnamefont {A.}~\bibnamefont
  {Kitaev}},\ }\href {https://doi.org/10.1070/1063-7869/44/10S/S29} {\bibfield
  {journal} {\bibinfo  {journal} {Phys. Usp.}\ }\textbf {\bibinfo {volume}
  {44}},\ \bibinfo {pages} {131} (\bibinfo {year} {2001})}\BibitemShut
  {NoStop}%
\bibitem [{\citenamefont {Wen}(2003)}]{wen2003b}%
  \BibitemOpen
  \bibfield  {author} {\bibinfo {author} {\bibfnamefont {X.-G.}\ \bibnamefont
  {Wen}},\ }\href {https://doi.org/10.1103/PhysRevD.68.065003} {\bibfield
  {journal} {\bibinfo  {journal} {Phys. Rev. D}\ }\textbf {\bibinfo {volume}
  {68}},\ \bibinfo {pages} {065003} (\bibinfo {year} {2003})}\BibitemShut
  {NoStop}%
\bibitem [{\citenamefont {Yao}\ \emph {et~al.}(2009)\citenamefont {Yao},
  \citenamefont {Zhang},\ and\ \citenamefont {Kivelson}}]{yao2009}%
  \BibitemOpen
  \bibfield  {author} {\bibinfo {author} {\bibfnamefont {H.}~\bibnamefont
  {Yao}}, \bibinfo {author} {\bibfnamefont {S.-C.}\ \bibnamefont {Zhang}},\
  and\ \bibinfo {author} {\bibfnamefont {S.~A.}\ \bibnamefont {Kivelson}},\
  }\href {https://doi.org/10.1103/PhysRevLett.102.217202} {\bibfield  {journal}
  {\bibinfo  {journal} {Phys. Rev. Lett.}\ }\textbf {\bibinfo {volume} {102}},\
  \bibinfo {pages} {217202} (\bibinfo {year} {2009})}\BibitemShut {NoStop}%
\bibitem [{\citenamefont {Wu}\ \emph {et~al.}(2009)\citenamefont {Wu},
  \citenamefont {Arovas},\ and\ \citenamefont {Hung}}]{wu2009}%
  \BibitemOpen
  \bibfield  {author} {\bibinfo {author} {\bibfnamefont {C.}~\bibnamefont
  {Wu}}, \bibinfo {author} {\bibfnamefont {D.}~\bibnamefont {Arovas}},\ and\
  \bibinfo {author} {\bibfnamefont {H.-H.}\ \bibnamefont {Hung}},\ }\href
  {https://doi.org/10.1103/PhysRevB.79.134427} {\bibfield  {journal} {\bibinfo
  {journal} {Phys. Rev. B}\ }\textbf {\bibinfo {volume} {79}},\ \bibinfo
  {pages} {134427} (\bibinfo {year} {2009})}\BibitemShut {NoStop}%
\bibitem [{\citenamefont {Ryu}(2009)}]{ryu2009}%
  \BibitemOpen
  \bibfield  {author} {\bibinfo {author} {\bibfnamefont {S.}~\bibnamefont
  {Ryu}},\ }\href {https://doi.org/10.1103/PhysRevB.79.075124} {\bibfield
  {journal} {\bibinfo  {journal} {Phys. Rev. B}\ }\textbf {\bibinfo {volume}
  {79}},\ \bibinfo {pages} {075124} (\bibinfo {year} {2009})}\BibitemShut
  {NoStop}%
\bibitem [{\citenamefont {Seifert}\ \emph {et~al.}(2020)\citenamefont
  {Seifert}, \citenamefont {Dong}, \citenamefont {Chulliparambil},
  \citenamefont {Vojta}, \citenamefont {Tu},\ and\ \citenamefont
  {Janssen}}]{seifert2020}%
  \BibitemOpen
  \bibfield  {author} {\bibinfo {author} {\bibfnamefont {U.~F.~P.}\
  \bibnamefont {Seifert}}, \bibinfo {author} {\bibfnamefont {X.-Y.}\
  \bibnamefont {Dong}}, \bibinfo {author} {\bibfnamefont {S.}~\bibnamefont
  {Chulliparambil}}, \bibinfo {author} {\bibfnamefont {M.}~\bibnamefont
  {Vojta}}, \bibinfo {author} {\bibfnamefont {H.-H.}\ \bibnamefont {Tu}},\ and\
  \bibinfo {author} {\bibfnamefont {L.}~\bibnamefont {Janssen}},\ }\href
  {https://doi.org/10.1103/PhysRevLett.125.257202} {\bibfield  {journal}
  {\bibinfo  {journal} {Phys. Rev. Lett.}\ }\textbf {\bibinfo {volume} {125}},\
  \bibinfo {pages} {257202} (\bibinfo {year} {2020})}\BibitemShut {NoStop}%
\bibitem [{\citenamefont {Chulliparambil}\ \emph {et~al.}(2021)\citenamefont
  {Chulliparambil}, \citenamefont {Janssen}, \citenamefont {Vojta},
  \citenamefont {Tu},\ and\ \citenamefont {Seifert}}]{chulliparambil2021}%
  \BibitemOpen
  \bibfield  {author} {\bibinfo {author} {\bibfnamefont {S.}~\bibnamefont
  {Chulliparambil}}, \bibinfo {author} {\bibfnamefont {L.}~\bibnamefont
  {Janssen}}, \bibinfo {author} {\bibfnamefont {M.}~\bibnamefont {Vojta}},
  \bibinfo {author} {\bibfnamefont {H.-H.}\ \bibnamefont {Tu}},\ and\ \bibinfo
  {author} {\bibfnamefont {U.~F.~P.}\ \bibnamefont {Seifert}},\ }\href
  {https://doi.org/10.1103/PhysRevB.103.075144} {\bibfield  {journal} {\bibinfo
   {journal} {Phys. Rev. B}\ }\textbf {\bibinfo {volume} {103}},\ \bibinfo
  {pages} {075144} (\bibinfo {year} {2021})}\BibitemShut {NoStop}%
\bibitem [{\citenamefont {Jin}\ \emph {et~al.}(2022)\citenamefont {Jin},
  \citenamefont {Natori}, \citenamefont {Pollmann},\ and\ \citenamefont
  {Knolle}}]{jin2022}%
  \BibitemOpen
  \bibfield  {author} {\bibinfo {author} {\bibfnamefont {H.-K.}\ \bibnamefont
  {Jin}}, \bibinfo {author} {\bibfnamefont {W.~M.~H.}\ \bibnamefont {Natori}},
  \bibinfo {author} {\bibfnamefont {F.}~\bibnamefont {Pollmann}},\ and\
  \bibinfo {author} {\bibfnamefont {J.}~\bibnamefont {Knolle}},\ }\href
  {https://doi.org/10.1038/s41467-022-31503-0} {\bibfield  {journal} {\bibinfo
  {journal} {Nat. Commun.}\ }\textbf {\bibinfo {volume} {13}},\ \bibinfo
  {pages} {3813} (\bibinfo {year} {2022})}\BibitemShut {NoStop}%
\bibitem [{\citenamefont {Natori}\ \emph {et~al.}(2023)\citenamefont {Natori},
  \citenamefont {Jin},\ and\ \citenamefont {Knolle}}]{natori2023}%
  \BibitemOpen
  \bibfield  {author} {\bibinfo {author} {\bibfnamefont {W.~M.~H.}\
  \bibnamefont {Natori}}, \bibinfo {author} {\bibfnamefont {H.-K.}\
  \bibnamefont {Jin}},\ and\ \bibinfo {author} {\bibfnamefont {J.}~\bibnamefont
  {Knolle}},\ }\href {https://doi.org/10.1103/PhysRevB.108.075111} {\bibfield
  {journal} {\bibinfo  {journal} {Phys. Rev. B}\ }\textbf {\bibinfo {volume}
  {108}},\ \bibinfo {pages} {075111} (\bibinfo {year} {2023})}\BibitemShut
  {NoStop}%
\bibitem [{Note1()}]{Note1}%
  \BibitemOpen
  \bibinfo {note} {A caveat is in order. Given the close similarity between
  this Hamiltonian and that of the TFI chain~\protect \eqref
  {eq:TFIM-hamiltonian}, it is tempting to interpret the coefficient of the
  $h$-term in Eq.~\protect \eqref {eq:son-hamiltonian} as an external magnetic
  field. However, it is generally not the case except when $n=1$, for which the
  model reduces to the TFI chain itself; an example is given by the model with
  $n=2$, as detailed in Appendix~\ref {append:clifford-rep}. Instead, it is
  better to view the $J$- and $h$-terms as inter- and intra-site couplings,
  respectively. Moreover, the Hamiltonian~\protect \eqref {eq:son-hamiltonian}
  exhibits the $\protect \mathrm {SO}(n)$ symmetry for generic values of $J$
  and $h$; this is not to be confused with the Hamiltonian given in~\protect
  \eqref {eq:parent-hamiltonian-enhanced-symmetry}, which is another parent
  Hamiltonian of the ground state(s) when $h=0$ and is $\protect \mathrm
  {SO}(n+1)$-symmetric.}\BibitemShut {Stop}%
\bibitem [{\citenamefont {Mumford}(2006)}]{mumford2006}%
  \BibitemOpen
  \bibfield  {author} {\bibinfo {author} {\bibfnamefont {D.}~\bibnamefont
  {Mumford}},\ }\href@noop {} {\emph {\bibinfo {title} {Tata Lectures on Theta
  I}}}\ (\bibinfo  {publisher} {Birkhäuser, Boston, MA},\ \bibinfo {year}
  {2006})\BibitemShut {NoStop}%
\bibitem [{\citenamefont {Moore}\ and\ \citenamefont
  {Seiberg}(1988)}]{moore1988}%
  \BibitemOpen
  \bibfield  {author} {\bibinfo {author} {\bibfnamefont {G.}~\bibnamefont
  {Moore}}\ and\ \bibinfo {author} {\bibfnamefont {N.}~\bibnamefont
  {Seiberg}},\ }\href {https://doi.org/10.1016/0370-2693(88)91796-0} {\bibfield
   {journal} {\bibinfo  {journal} {Phys. Lett. B}\ }\textbf {\bibinfo {volume}
  {212}},\ \bibinfo {pages} {451} (\bibinfo {year} {1988})}\BibitemShut
  {NoStop}%
\bibitem [{\citenamefont {Cardy}(1989)}]{cardy1989}%
  \BibitemOpen
  \bibfield  {author} {\bibinfo {author} {\bibfnamefont {J.~L.}\ \bibnamefont
  {Cardy}},\ }\href {https://doi.org/10.1016/0550-3213(89)90521-X} {\bibfield
  {journal} {\bibinfo  {journal} {Nucl. Phys. B}\ }\textbf {\bibinfo {volume}
  {324}},\ \bibinfo {pages} {581} (\bibinfo {year} {1989})}\BibitemShut
  {NoStop}%
\bibitem [{\citenamefont {Belavin}\ \emph {et~al.}(1984)\citenamefont
  {Belavin}, \citenamefont {Polyakov},\ and\ \citenamefont
  {Zamolodchikov}}]{belavin1984}%
  \BibitemOpen
  \bibfield  {author} {\bibinfo {author} {\bibfnamefont {A.}~\bibnamefont
  {Belavin}}, \bibinfo {author} {\bibfnamefont {A.}~\bibnamefont {Polyakov}},\
  and\ \bibinfo {author} {\bibfnamefont {A.}~\bibnamefont {Zamolodchikov}},\
  }\href {https://doi.org/10.1016/0550-3213(84)90052-X} {\bibfield  {journal}
  {\bibinfo  {journal} {Nucl. Phys. B}\ }\textbf {\bibinfo {volume} {241}},\
  \bibinfo {pages} {333} (\bibinfo {year} {1984})}\BibitemShut {NoStop}%
\bibitem [{\citenamefont {Friedan}\ \emph {et~al.}(1984)\citenamefont
  {Friedan}, \citenamefont {Qiu},\ and\ \citenamefont {Shenker}}]{friedan1984}%
  \BibitemOpen
  \bibfield  {author} {\bibinfo {author} {\bibfnamefont {D.}~\bibnamefont
  {Friedan}}, \bibinfo {author} {\bibfnamefont {Z.}~\bibnamefont {Qiu}},\ and\
  \bibinfo {author} {\bibfnamefont {S.}~\bibnamefont {Shenker}},\ }\href
  {https://doi.org/10.1103/PhysRevLett.52.1575} {\bibfield  {journal} {\bibinfo
   {journal} {Phys. Rev. Lett.}\ }\textbf {\bibinfo {volume} {52}},\ \bibinfo
  {pages} {1575} (\bibinfo {year} {1984})}\BibitemShut {NoStop}%
\bibitem [{\citenamefont {Witten}(1984)}]{witten1984}%
  \BibitemOpen
  \bibfield  {author} {\bibinfo {author} {\bibfnamefont {E.}~\bibnamefont
  {Witten}},\ }\href {https://doi.org/10.1007/BF01215276} {\bibfield  {journal}
  {\bibinfo  {journal} {Commun. Math. Phys.}\ }\textbf {\bibinfo {volume}
  {92}},\ \bibinfo {pages} {455} (\bibinfo {year} {1984})}\BibitemShut
  {NoStop}%
\bibitem [{Note2()}]{Note2}%
  \BibitemOpen
  \bibinfo {note} {Note that there are different notations for this theory in
  the literature. For example, the same theory is referred to as the $\protect
  \mathrm {SO}(n)_{1}$ WZW model in Ref.~\cite {francesco1997}; here, we follow
  the terminology of Ref.~\cite {seiberg2016} where the name $\protect \mathrm
  {SO}(n)_{1}$ is reserved for the corresponding ``spin CFT'', which depends on
  a particular choice of spin structures (i.e., boundary conditions of the
  fermions along the cycles of the spacetime torus). The $\protect \mathrm
  {Spin}(n)_{1}$ CFT, on the contrary, is non-spin as the spin structures are
  summed over in computing the partition function}\BibitemShut {NoStop}%
\bibitem [{\citenamefont {Francesco}\ \emph {et~al.}(1997)\citenamefont
  {Francesco}, \citenamefont {Mathieu},\ and\ \citenamefont
  {S\'en\'echal}}]{francesco1997}%
  \BibitemOpen
  \bibfield  {author} {\bibinfo {author} {\bibfnamefont {P.~D.}\ \bibnamefont
  {Francesco}}, \bibinfo {author} {\bibfnamefont {P.}~\bibnamefont {Mathieu}},\
  and\ \bibinfo {author} {\bibfnamefont {D.}~\bibnamefont {S\'en\'echal}},\
  }\href@noop {} {\emph {\bibinfo {title} {Conformal Field Theory}}}\ (\bibinfo
   {publisher} {Springer-Verlag, New York},\ \bibinfo {year}
  {1997})\BibitemShut {NoStop}%
\bibitem [{\citenamefont {Zhang}\ \emph {et~al.}(2022)\citenamefont {Zhang},
  \citenamefont {Wu}, \citenamefont {Xiang},\ and\ \citenamefont
  {Tu}}]{zhang2022}%
  \BibitemOpen
  \bibfield  {author} {\bibinfo {author} {\bibfnamefont {H.-C.}\ \bibnamefont
  {Zhang}}, \bibinfo {author} {\bibfnamefont {Y.-H.}\ \bibnamefont {Wu}},
  \bibinfo {author} {\bibfnamefont {T.}~\bibnamefont {Xiang}},\ and\ \bibinfo
  {author} {\bibfnamefont {H.-H.}\ \bibnamefont {Tu}},\ }\href
  {https://doi.org/10.1016/j.nuclphysb.2022.115712} {\bibfield  {journal}
  {\bibinfo  {journal} {Nucl. Phys. B}\ }\textbf {\bibinfo {volume} {976}},\
  \bibinfo {pages} {115712} (\bibinfo {year} {2022})}\BibitemShut {NoStop}%
\bibitem [{\citenamefont {M\aa{}nsson}\ \emph {et~al.}(2013)\citenamefont
  {M\aa{}nsson}, \citenamefont {Lahtinen}, \citenamefont {Suorsa},\ and\
  \citenamefont {Ardonne}}]{maansson2013}%
  \BibitemOpen
  \bibfield  {author} {\bibinfo {author} {\bibfnamefont {T.}~\bibnamefont
  {M\aa{}nsson}}, \bibinfo {author} {\bibfnamefont {V.}~\bibnamefont
  {Lahtinen}}, \bibinfo {author} {\bibfnamefont {J.}~\bibnamefont {Suorsa}},\
  and\ \bibinfo {author} {\bibfnamefont {E.}~\bibnamefont {Ardonne}},\ }\href
  {https://doi.org/10.1103/PhysRevB.88.041403} {\bibfield  {journal} {\bibinfo
  {journal} {Phys. Rev. B}\ }\textbf {\bibinfo {volume} {88}},\ \bibinfo
  {pages} {041403(R)} (\bibinfo {year} {2013})}\BibitemShut {NoStop}%
\bibitem [{\citenamefont {Lahtinen}\ \emph {et~al.}(2014)\citenamefont
  {Lahtinen}, \citenamefont {M\aa{}nsson},\ and\ \citenamefont
  {Ardonne}}]{lahtinen2014}%
  \BibitemOpen
  \bibfield  {author} {\bibinfo {author} {\bibfnamefont {V.}~\bibnamefont
  {Lahtinen}}, \bibinfo {author} {\bibfnamefont {T.}~\bibnamefont
  {M\aa{}nsson}},\ and\ \bibinfo {author} {\bibfnamefont {E.}~\bibnamefont
  {Ardonne}},\ }\href {https://doi.org/10.1103/PhysRevB.89.014409} {\bibfield
  {journal} {\bibinfo  {journal} {Phys. Rev. B}\ }\textbf {\bibinfo {volume}
  {89}},\ \bibinfo {pages} {014409} (\bibinfo {year} {2014})}\BibitemShut
  {NoStop}%
\bibitem [{\citenamefont {Lahtinen}\ and\ \citenamefont
  {Ardonne}(2015)}]{lahtinen2015}%
  \BibitemOpen
  \bibfield  {author} {\bibinfo {author} {\bibfnamefont {V.}~\bibnamefont
  {Lahtinen}}\ and\ \bibinfo {author} {\bibfnamefont {E.}~\bibnamefont
  {Ardonne}},\ }\href {https://doi.org/10.1103/PhysRevLett.115.237203}
  {\bibfield  {journal} {\bibinfo  {journal} {Phys. Rev. Lett.}\ }\textbf
  {\bibinfo {volume} {115}},\ \bibinfo {pages} {237203} (\bibinfo {year}
  {2015})}\BibitemShut {NoStop}%
\bibitem [{\citenamefont {Reshetikhin}(1983)}]{reshetikhin1983}%
  \BibitemOpen
  \bibfield  {author} {\bibinfo {author} {\bibfnamefont {N.~Y.}\ \bibnamefont
  {Reshetikhin}},\ }\href {https://doi.org/10.1007/BF00400435} {\bibfield
  {journal} {\bibinfo  {journal} {Lett. Math. Phys.}\ }\textbf {\bibinfo
  {volume} {7}},\ \bibinfo {pages} {205} (\bibinfo {year} {1983})}\BibitemShut
  {NoStop}%
\bibitem [{\citenamefont {Reshetikhin}(1985)}]{reshetikhin1985}%
  \BibitemOpen
  \bibfield  {author} {\bibinfo {author} {\bibfnamefont {N.~Y.}\ \bibnamefont
  {Reshetikhin}},\ }\href {https://doi.org/10.1007/BF01017501} {\bibfield
  {journal} {\bibinfo  {journal} {Theor. Math. Phys.}\ }\textbf {\bibinfo
  {volume} {63}},\ \bibinfo {pages} {555} (\bibinfo {year} {1985})}\BibitemShut
  {NoStop}%
\bibitem [{\citenamefont {Tu}(2013)}]{tu2013a}%
  \BibitemOpen
  \bibfield  {author} {\bibinfo {author} {\bibfnamefont {H.-H.}\ \bibnamefont
  {Tu}},\ }\href {https://doi.org/10.1103/PhysRevB.87.041103} {\bibfield
  {journal} {\bibinfo  {journal} {Phys. Rev. B}\ }\textbf {\bibinfo {volume}
  {87}},\ \bibinfo {pages} {041103(R)} (\bibinfo {year} {2013})}\BibitemShut
  {NoStop}%
\bibitem [{\citenamefont {Haldane}(1988)}]{haldane1988}%
  \BibitemOpen
  \bibfield  {author} {\bibinfo {author} {\bibfnamefont {F.~D.~M.}\
  \bibnamefont {Haldane}},\ }\href {https://doi.org/10.1103/PhysRevLett.60.635}
  {\bibfield  {journal} {\bibinfo  {journal} {Phys. Rev. Lett.}\ }\textbf
  {\bibinfo {volume} {60}},\ \bibinfo {pages} {635} (\bibinfo {year}
  {1988})}\BibitemShut {NoStop}%
\bibitem [{\citenamefont {Shastry}(1988)}]{shastry1988}%
  \BibitemOpen
  \bibfield  {author} {\bibinfo {author} {\bibfnamefont {B.~S.}\ \bibnamefont
  {Shastry}},\ }\href {https://doi.org/10.1103/PhysRevLett.60.639} {\bibfield
  {journal} {\bibinfo  {journal} {Phys. Rev. Lett.}\ }\textbf {\bibinfo
  {volume} {60}},\ \bibinfo {pages} {639} (\bibinfo {year} {1988})}\BibitemShut
  {NoStop}%
\bibitem [{\citenamefont {Fendley}(2012)}]{fendley2012}%
  \BibitemOpen
  \bibfield  {author} {\bibinfo {author} {\bibfnamefont {P.}~\bibnamefont
  {Fendley}},\ }\href {https://doi.org/10.1088/1742-5468/2012/11/P11020}
  {\bibfield  {journal} {\bibinfo  {journal} {J. Stat. Mech.}\ }\textbf
  {\bibinfo {volume} {2012}},\ \bibinfo {pages} {P11020} (\bibinfo {year}
  {2012})}\BibitemShut {NoStop}%
\bibitem [{Note3()}]{Note3}%
  \BibitemOpen
  \bibinfo {note} {To tally with the notation we used in Sec.~\ref
  {sec:criticality}, we note that the representation $D^{k}$ corresponds to the
  primary field $\protect \bm {s}$ of the $\protect \mathrm {Spin}(2k+1)_1$
  CFT, and $D^{k,\pm }$ correspond to $\protect \bm {s}_{\pm }$ of the
  $\protect \mathrm {Spin}(2k)_1$ CFT}\BibitemShut {NoStop}%
\bibitem [{\citenamefont {Georgi}(1999)}]{georgi1999}%
  \BibitemOpen
  \bibfield  {author} {\bibinfo {author} {\bibfnamefont {H.}~\bibnamefont
  {Georgi}},\ }\href@noop {} {\emph {\bibinfo {title} {Lie Algebras in Particle
  Physics}}}\ (\bibinfo  {publisher} {Perseus Books, Reading, MA},\ \bibinfo
  {year} {1999})\BibitemShut {NoStop}%
\bibitem [{\citenamefont {Gu}\ and\ \citenamefont {Wen}(2009)}]{gu2009}%
  \BibitemOpen
  \bibfield  {author} {\bibinfo {author} {\bibfnamefont {Z.-C.}\ \bibnamefont
  {Gu}}\ and\ \bibinfo {author} {\bibfnamefont {X.-G.}\ \bibnamefont {Wen}},\
  }\href {https://doi.org/10.1103/PhysRevB.80.155131} {\bibfield  {journal}
  {\bibinfo  {journal} {Phys. Rev. B}\ }\textbf {\bibinfo {volume} {80}},\
  \bibinfo {pages} {155131} (\bibinfo {year} {2009})}\BibitemShut {NoStop}%
\bibitem [{\citenamefont {Pollmann}\ \emph {et~al.}(2010)\citenamefont
  {Pollmann}, \citenamefont {Turner}, \citenamefont {Berg},\ and\ \citenamefont
  {Oshikawa}}]{pollmann2010}%
  \BibitemOpen
  \bibfield  {author} {\bibinfo {author} {\bibfnamefont {F.}~\bibnamefont
  {Pollmann}}, \bibinfo {author} {\bibfnamefont {A.~M.}\ \bibnamefont
  {Turner}}, \bibinfo {author} {\bibfnamefont {E.}~\bibnamefont {Berg}},\ and\
  \bibinfo {author} {\bibfnamefont {M.}~\bibnamefont {Oshikawa}},\ }\href
  {https://doi.org/10.1103/PhysRevB.81.064439} {\bibfield  {journal} {\bibinfo
  {journal} {Phys. Rev. B}\ }\textbf {\bibinfo {volume} {81}},\ \bibinfo
  {pages} {064439} (\bibinfo {year} {2010})}\BibitemShut {NoStop}%
\bibitem [{\citenamefont {Chen}\ \emph {et~al.}(2011)\citenamefont {Chen},
  \citenamefont {Gu},\ and\ \citenamefont {Wen}}]{chen2011}%
  \BibitemOpen
  \bibfield  {author} {\bibinfo {author} {\bibfnamefont {X.}~\bibnamefont
  {Chen}}, \bibinfo {author} {\bibfnamefont {Z.-C.}\ \bibnamefont {Gu}},\ and\
  \bibinfo {author} {\bibfnamefont {X.-G.}\ \bibnamefont {Wen}},\ }\href
  {https://doi.org/10.1103/PhysRevB.83.035107} {\bibfield  {journal} {\bibinfo
  {journal} {Phys. Rev. B}\ }\textbf {\bibinfo {volume} {83}},\ \bibinfo
  {pages} {035107} (\bibinfo {year} {2011})}\BibitemShut {NoStop}%
\bibitem [{\citenamefont {Schuch}\ \emph {et~al.}(2011)\citenamefont {Schuch},
  \citenamefont {P\'erez-Garc\'{\i}a},\ and\ \citenamefont
  {Cirac}}]{schuch2011}%
  \BibitemOpen
  \bibfield  {author} {\bibinfo {author} {\bibfnamefont {N.}~\bibnamefont
  {Schuch}}, \bibinfo {author} {\bibfnamefont {D.}~\bibnamefont
  {P\'erez-Garc\'{\i}a}},\ and\ \bibinfo {author} {\bibfnamefont
  {I.}~\bibnamefont {Cirac}},\ }\href
  {https://doi.org/10.1103/PhysRevB.84.165139} {\bibfield  {journal} {\bibinfo
  {journal} {Phys. Rev. B}\ }\textbf {\bibinfo {volume} {84}},\ \bibinfo
  {pages} {165139} (\bibinfo {year} {2011})}\BibitemShut {NoStop}%
\bibitem [{\citenamefont {Duivenvoorden}\ and\ \citenamefont
  {Quella}(2013)}]{duivenvoorden2013}%
  \BibitemOpen
  \bibfield  {author} {\bibinfo {author} {\bibfnamefont {K.}~\bibnamefont
  {Duivenvoorden}}\ and\ \bibinfo {author} {\bibfnamefont {T.}~\bibnamefont
  {Quella}},\ }\href {https://doi.org/10.1103/PhysRevB.87.125145} {\bibfield
  {journal} {\bibinfo  {journal} {Phys. Rev. B}\ }\textbf {\bibinfo {volume}
  {87}},\ \bibinfo {pages} {125145} (\bibinfo {year} {2013})}\BibitemShut
  {NoStop}%
\bibitem [{\citenamefont {Ramond}(2010)}]{ramond2010}%
  \BibitemOpen
  \bibfield  {author} {\bibinfo {author} {\bibfnamefont {P.}~\bibnamefont
  {Ramond}},\ }\href@noop {} {\emph {\bibinfo {title} {Group Theory: A
  Physicist’s Survey}}}\ (\bibinfo  {publisher} {Cambridge University Press,
  New York},\ \bibinfo {year} {2010})\BibitemShut {NoStop}%
\bibitem [{\citenamefont {Temperley}\ and\ \citenamefont
  {Lieb}(1971)}]{temperley1971}%
  \BibitemOpen
  \bibfield  {author} {\bibinfo {author} {\bibfnamefont {H.}~\bibnamefont
  {Temperley}}\ and\ \bibinfo {author} {\bibfnamefont {E.~H.}\ \bibnamefont
  {Lieb}},\ }\href {https://doi.org/10.1098/rspa.1971.0067} {\bibfield
  {journal} {\bibinfo  {journal} {Proc. R. Soc. A: Math. Phys. Eng. Sci.}\
  }\textbf {\bibinfo {volume} {322}},\ \bibinfo {pages} {251} (\bibinfo {year}
  {1971})}\BibitemShut {NoStop}%
\bibitem [{\citenamefont {Levy}(1991)}]{levy1991}%
  \BibitemOpen
  \bibfield  {author} {\bibinfo {author} {\bibfnamefont {D.}~\bibnamefont
  {Levy}},\ }\href {https://doi.org/10.1103/PhysRevLett.67.1971} {\bibfield
  {journal} {\bibinfo  {journal} {Phys. Rev. Lett.}\ }\textbf {\bibinfo
  {volume} {67}},\ \bibinfo {pages} {1971} (\bibinfo {year}
  {1991})}\BibitemShut {NoStop}%
\bibitem [{\citenamefont {Montes}\ \emph {et~al.}(2017)\citenamefont {Montes},
  \citenamefont {Rodr\'{\i}guez-Laguna}, \citenamefont {Tu},\ and\
  \citenamefont {Sierra}}]{montes2017}%
  \BibitemOpen
  \bibfield  {author} {\bibinfo {author} {\bibfnamefont {S.}~\bibnamefont
  {Montes}}, \bibinfo {author} {\bibfnamefont {J.}~\bibnamefont
  {Rodr\'{\i}guez-Laguna}}, \bibinfo {author} {\bibfnamefont {H.-H.}\
  \bibnamefont {Tu}},\ and\ \bibinfo {author} {\bibfnamefont {G.}~\bibnamefont
  {Sierra}},\ }\href {https://doi.org/10.1103/PhysRevB.95.085146} {\bibfield
  {journal} {\bibinfo  {journal} {Phys. Rev. B}\ }\textbf {\bibinfo {volume}
  {95}},\ \bibinfo {pages} {085146} (\bibinfo {year} {2017})}\BibitemShut
  {NoStop}%
\bibitem [{\citenamefont {Davies}(1990)}]{davies1990}%
  \BibitemOpen
  \bibfield  {author} {\bibinfo {author} {\bibfnamefont {B.}~\bibnamefont
  {Davies}},\ }\href {https://doi.org/10.1088/0305-4470/23/12/010} {\bibfield
  {journal} {\bibinfo  {journal} {J. Phys. A: Math. Gen.}\ }\textbf {\bibinfo
  {volume} {23}},\ \bibinfo {pages} {2245} (\bibinfo {year}
  {1990})}\BibitemShut {NoStop}%
\bibitem [{\citenamefont {Davies}(1991)}]{davies1991}%
  \BibitemOpen
  \bibfield  {author} {\bibinfo {author} {\bibfnamefont {B.}~\bibnamefont
  {Davies}},\ }\href {https://doi.org/10.1063/1.529036} {\bibfield  {journal}
  {\bibinfo  {journal} {J. Math. Phys.}\ }\textbf {\bibinfo {volume} {32}},\
  \bibinfo {pages} {2945} (\bibinfo {year} {1991})}\BibitemShut {NoStop}%
\bibitem [{\citenamefont {Perk}()}]{perk2017}%
  \BibitemOpen
  \bibfield  {author} {\bibinfo {author} {\bibfnamefont {J.~H.~H.}\
  \bibnamefont {Perk}},\ }\href@noop {} {\ }\Eprint
  {https://arxiv.org/abs/1710.03384} {arXiv:1710.03384} \BibitemShut {NoStop}%
\bibitem [{\citenamefont {Miao}(2022)}]{miao2022}%
  \BibitemOpen
  \bibfield  {author} {\bibinfo {author} {\bibfnamefont {Y.}~\bibnamefont
  {Miao}},\ }\href {https://doi.org/10.21468/SciPostPhys.13.3.070} {\bibfield
  {journal} {\bibinfo  {journal} {SciPost Phys.}\ }\textbf {\bibinfo {volume}
  {13}},\ \bibinfo {pages} {070} (\bibinfo {year} {2022})}\BibitemShut
  {NoStop}%
\bibitem [{\citenamefont {Li}\ \emph {et~al.}(2022)\citenamefont {Li},
  \citenamefont {Quito}, \citenamefont {Miranda}, \citenamefont {Pereira},
  \citenamefont {Affleck},\ and\ \citenamefont {Lopes}}]{li2022}%
  \BibitemOpen
  \bibfield  {author} {\bibinfo {author} {\bibfnamefont {C.}~\bibnamefont
  {Li}}, \bibinfo {author} {\bibfnamefont {V.~L.}\ \bibnamefont {Quito}},
  \bibinfo {author} {\bibfnamefont {E.}~\bibnamefont {Miranda}}, \bibinfo
  {author} {\bibfnamefont {R.}~\bibnamefont {Pereira}}, \bibinfo {author}
  {\bibfnamefont {I.}~\bibnamefont {Affleck}},\ and\ \bibinfo {author}
  {\bibfnamefont {P.~L.~S.}\ \bibnamefont {Lopes}},\ }\href
  {https://doi.org/10.1103/PhysRevB.105.085140} {\bibfield  {journal} {\bibinfo
   {journal} {Phys. Rev. B}\ }\textbf {\bibinfo {volume} {105}},\ \bibinfo
  {pages} {085140} (\bibinfo {year} {2022})}\BibitemShut {NoStop}%
\bibitem [{\citenamefont {Hu}\ and\ \citenamefont {Kane}(2018)}]{hu2018}%
  \BibitemOpen
  \bibfield  {author} {\bibinfo {author} {\bibfnamefont {Y.}~\bibnamefont
  {Hu}}\ and\ \bibinfo {author} {\bibfnamefont {C.~L.}\ \bibnamefont {Kane}},\
  }\href {https://doi.org/10.1103/PhysRevLett.120.066801} {\bibfield  {journal}
  {\bibinfo  {journal} {Phys. Rev. Lett.}\ }\textbf {\bibinfo {volume} {120}},\
  \bibinfo {pages} {066801} (\bibinfo {year} {2018})}\BibitemShut {NoStop}%
\bibitem [{\citenamefont {Verresen}\ \emph {et~al.}(2018)\citenamefont
  {Verresen}, \citenamefont {Jones},\ and\ \citenamefont
  {Pollmann}}]{verresen2018}%
  \BibitemOpen
  \bibfield  {author} {\bibinfo {author} {\bibfnamefont {R.}~\bibnamefont
  {Verresen}}, \bibinfo {author} {\bibfnamefont {N.~G.}\ \bibnamefont
  {Jones}},\ and\ \bibinfo {author} {\bibfnamefont {F.}~\bibnamefont
  {Pollmann}},\ }\href {https://doi.org/10.1103/PhysRevLett.120.057001}
  {\bibfield  {journal} {\bibinfo  {journal} {Phys. Rev. Lett.}\ }\textbf
  {\bibinfo {volume} {120}},\ \bibinfo {pages} {057001} (\bibinfo {year}
  {2018})}\BibitemShut {NoStop}%
\bibitem [{\citenamefont {Parker}\ \emph {et~al.}(2018)\citenamefont {Parker},
  \citenamefont {Scaffidi},\ and\ \citenamefont {Vasseur}}]{parker2018}%
  \BibitemOpen
  \bibfield  {author} {\bibinfo {author} {\bibfnamefont {D.~E.}\ \bibnamefont
  {Parker}}, \bibinfo {author} {\bibfnamefont {T.}~\bibnamefont {Scaffidi}},\
  and\ \bibinfo {author} {\bibfnamefont {R.}~\bibnamefont {Vasseur}},\ }\href
  {https://doi.org/10.1103/PhysRevB.97.165114} {\bibfield  {journal} {\bibinfo
  {journal} {Phys. Rev. B}\ }\textbf {\bibinfo {volume} {97}},\ \bibinfo
  {pages} {165114} (\bibinfo {year} {2018})}\BibitemShut {NoStop}%
\bibitem [{\citenamefont {Jones}\ and\ \citenamefont
  {Verresen}(2019)}]{jones2019}%
  \BibitemOpen
  \bibfield  {author} {\bibinfo {author} {\bibfnamefont {N.~G.}\ \bibnamefont
  {Jones}}\ and\ \bibinfo {author} {\bibfnamefont {R.}~\bibnamefont
  {Verresen}},\ }\href {https://doi.org/10.1007/s10955-019-02257-9} {\bibfield
  {journal} {\bibinfo  {journal} {J. Stat. Phys.}\ }\textbf {\bibinfo {volume}
  {175}},\ \bibinfo {pages} {1164} (\bibinfo {year} {2019})}\BibitemShut
  {NoStop}%
\bibitem [{\citenamefont {Verresen}\ \emph {et~al.}(2021)\citenamefont
  {Verresen}, \citenamefont {Thorngren}, \citenamefont {Jones},\ and\
  \citenamefont {Pollmann}}]{verresen2021}%
  \BibitemOpen
  \bibfield  {author} {\bibinfo {author} {\bibfnamefont {R.}~\bibnamefont
  {Verresen}}, \bibinfo {author} {\bibfnamefont {R.}~\bibnamefont {Thorngren}},
  \bibinfo {author} {\bibfnamefont {N.~G.}\ \bibnamefont {Jones}},\ and\
  \bibinfo {author} {\bibfnamefont {F.}~\bibnamefont {Pollmann}},\ }\href
  {https://doi.org/10.1103/PhysRevX.11.041059} {\bibfield  {journal} {\bibinfo
  {journal} {Phys. Rev. X}\ }\textbf {\bibinfo {volume} {11}},\ \bibinfo
  {pages} {041059} (\bibinfo {year} {2021})}\BibitemShut {NoStop}%
\bibitem [{\citenamefont {Jones}\ and\ \citenamefont
  {Linden}(2022)}]{jones2022}%
  \BibitemOpen
  \bibfield  {author} {\bibinfo {author} {\bibfnamefont {N.~G.}\ \bibnamefont
  {Jones}}\ and\ \bibinfo {author} {\bibfnamefont {N.}~\bibnamefont {Linden}},\
  }\href {https://doi.org/10.1063/5.0095870} {\bibfield  {journal} {\bibinfo
  {journal} {J. Math. Phys.}\ }\textbf {\bibinfo {volume} {63}},\ \bibinfo
  {pages} {101901} (\bibinfo {year} {2022})}\BibitemShut {NoStop}%
\bibitem [{\citenamefont {Yang}\ \emph {et~al.}(2023)\citenamefont {Yang},
  \citenamefont {Li}, \citenamefont {Okunishi},\ and\ \citenamefont
  {Katsura}}]{yang2023}%
  \BibitemOpen
  \bibfield  {author} {\bibinfo {author} {\bibfnamefont {H.}~\bibnamefont
  {Yang}}, \bibinfo {author} {\bibfnamefont {L.}~\bibnamefont {Li}}, \bibinfo
  {author} {\bibfnamefont {K.}~\bibnamefont {Okunishi}},\ and\ \bibinfo
  {author} {\bibfnamefont {H.}~\bibnamefont {Katsura}},\ }\href
  {https://doi.org/10.1103/PhysRevB.107.125158} {\bibfield  {journal} {\bibinfo
   {journal} {Phys. Rev. B}\ }\textbf {\bibinfo {volume} {107}},\ \bibinfo
  {pages} {125158} (\bibinfo {year} {2023})}\BibitemShut {NoStop}%
\bibitem [{\citenamefont {Li}\ \emph {et~al.}()\citenamefont {Li},
  \citenamefont {Oshikawa},\ and\ \citenamefont {Zheng}}]{li2023}%
  \BibitemOpen
  \bibfield  {author} {\bibinfo {author} {\bibfnamefont {L.}~\bibnamefont
  {Li}}, \bibinfo {author} {\bibfnamefont {M.}~\bibnamefont {Oshikawa}},\ and\
  \bibinfo {author} {\bibfnamefont {Y.}~\bibnamefont {Zheng}},\ }\href
  {https://arxiv.org/abs/1205.5989} {\ }\Eprint
  {https://arxiv.org/abs/2301.07899} {arXiv:2301.07899} \BibitemShut {NoStop}%
\bibitem [{\citenamefont {Yu}\ \emph {et~al.}(2022)\citenamefont {Yu},
  \citenamefont {Huang}, \citenamefont {Song}, \citenamefont {Xu},
  \citenamefont {Ding},\ and\ \citenamefont {Zhang}}]{yu2022}%
  \BibitemOpen
  \bibfield  {author} {\bibinfo {author} {\bibfnamefont {X.-J.}\ \bibnamefont
  {Yu}}, \bibinfo {author} {\bibfnamefont {R.-Z.}\ \bibnamefont {Huang}},
  \bibinfo {author} {\bibfnamefont {H.-H.}\ \bibnamefont {Song}}, \bibinfo
  {author} {\bibfnamefont {L.}~\bibnamefont {Xu}}, \bibinfo {author}
  {\bibfnamefont {C.}~\bibnamefont {Ding}},\ and\ \bibinfo {author}
  {\bibfnamefont {L.}~\bibnamefont {Zhang}},\ }\href
  {https://doi.org/10.1103/PhysRevLett.129.210601} {\bibfield  {journal}
  {\bibinfo  {journal} {Phys. Rev. Lett.}\ }\textbf {\bibinfo {volume} {129}},\
  \bibinfo {pages} {210601} (\bibinfo {year} {2022})}\BibitemShut {NoStop}%
\bibitem [{\citenamefont {Aasen}\ \emph {et~al.}(2016)\citenamefont {Aasen},
  \citenamefont {Mong},\ and\ \citenamefont {Fendley}}]{aasen2016}%
  \BibitemOpen
  \bibfield  {author} {\bibinfo {author} {\bibfnamefont {D.}~\bibnamefont
  {Aasen}}, \bibinfo {author} {\bibfnamefont {R.~S.~K.}\ \bibnamefont {Mong}},\
  and\ \bibinfo {author} {\bibfnamefont {P.}~\bibnamefont {Fendley}},\ }\href
  {https://doi.org/10.1088/1751-8113/49/35/354001} {\bibfield  {journal}
  {\bibinfo  {journal} {J. Phys. A: Math. Theor.}\ }\textbf {\bibinfo {volume}
  {49}},\ \bibinfo {pages} {354001} (\bibinfo {year} {2016})}\BibitemShut
  {NoStop}%
\bibitem [{\citenamefont {Aasen}\ \emph {et~al.}()\citenamefont {Aasen},
  \citenamefont {Fendley},\ and\ \citenamefont {Mong}}]{aasen2020}%
  \BibitemOpen
  \bibfield  {author} {\bibinfo {author} {\bibfnamefont {D.}~\bibnamefont
  {Aasen}}, \bibinfo {author} {\bibfnamefont {P.}~\bibnamefont {Fendley}},\
  and\ \bibinfo {author} {\bibfnamefont {R.~S.~K.}\ \bibnamefont {Mong}},\
  }\href {https://arxiv.org/abs/2008.08598} {\ }\Eprint
  {https://arxiv.org/abs/2008.08598} {arXiv:2008.08598} \BibitemShut {NoStop}%
\bibitem [{\citenamefont {Lootens}\ \emph {et~al.}(2023)\citenamefont
  {Lootens}, \citenamefont {Delcamp}, \citenamefont {Ortiz},\ and\
  \citenamefont {Verstraete}}]{lootens2021}%
  \BibitemOpen
  \bibfield  {author} {\bibinfo {author} {\bibfnamefont {L.}~\bibnamefont
  {Lootens}}, \bibinfo {author} {\bibfnamefont {C.}~\bibnamefont {Delcamp}},
  \bibinfo {author} {\bibfnamefont {G.}~\bibnamefont {Ortiz}},\ and\ \bibinfo
  {author} {\bibfnamefont {F.}~\bibnamefont {Verstraete}},\ }\href
  {https://doi.org/10.1103/PRXQuantum.4.020357} {\bibfield  {journal} {\bibinfo
   {journal} {PRX Quantum}\ }\textbf {\bibinfo {volume} {4}},\ \bibinfo {pages}
  {020357} (\bibinfo {year} {2023})}\BibitemShut {NoStop}%
\bibitem [{\citenamefont {Chugh}\ \emph {et~al.}(2022)\citenamefont {Chugh},
  \citenamefont {Dhochak}, \citenamefont {Divakaran}, \citenamefont {Narayan},\
  and\ \citenamefont {Pal}}]{chugh2022}%
  \BibitemOpen
  \bibfield  {author} {\bibinfo {author} {\bibfnamefont {Y.}~\bibnamefont
  {Chugh}}, \bibinfo {author} {\bibfnamefont {K.}~\bibnamefont {Dhochak}},
  \bibinfo {author} {\bibfnamefont {U.}~\bibnamefont {Divakaran}}, \bibinfo
  {author} {\bibfnamefont {P.}~\bibnamefont {Narayan}},\ and\ \bibinfo {author}
  {\bibfnamefont {A.~K.}\ \bibnamefont {Pal}},\ }\href
  {https://doi.org/10.1103/PhysRevE.106.024114} {\bibfield  {journal} {\bibinfo
   {journal} {Phys. Rev. E}\ }\textbf {\bibinfo {volume} {106}},\ \bibinfo
  {pages} {024114} (\bibinfo {year} {2022})}\BibitemShut {NoStop}%
\bibitem [{\citenamefont {Seiberg}\ and\ \citenamefont
  {Witten}()}]{seiberg2016}%
  \BibitemOpen
  \bibfield  {author} {\bibinfo {author} {\bibfnamefont {N.}~\bibnamefont
  {Seiberg}}\ and\ \bibinfo {author} {\bibfnamefont {E.}~\bibnamefont
  {Witten}},\ }\href {https://arxiv.org/abs/1602.04251} {}\Eprint
  {https://arxiv.org/abs/1602.04251} {arXiv:1602.04251} \BibitemShut {NoStop}%
\end{thebibliography}%

\end{document}